\documentclass[journal, onecolumn]{IEEEtran}
\usepackage{amsmath,amssymb}
\usepackage{graphicx}
\usepackage{epsfig}

\begin{document}
\title{Exploiting Spatial Interference Alignment and Opportunistic Scheduling in the Downlink of Interference Limited Systems}
\maketitle
\begin{center}
\begin{tabular}{c c }
 Kiran Kuchi \\
Indian Institute of Technology, Hyderabad, India\\
\end{tabular}
\end{center}

\vspace{1.0 cm}

\begin{abstract}
In this paper we analyze the performance of single stream and multi-stream spatial multiplexing (SM) systems employing opportunistic scheduling in the presence of interference. In the proposed downlink framework, every active user reports the post-processing signal-to-interference-plus-noise-power-ratio (post-SINR) or the receiver specific mutual information (MI) to its own transmitter using a feedback channel. The combination of scheduling and multi-antenna receiver processing leads to substantial interference suppression gain. Specifically, we show that opportunistic scheduling exploits \emph{spatial interference alignment} (SIA) property inherent to a multi-user system for effective interference mitigation. We obtain bounds for the outage probability and the sum outage capacity for single stream and multi stream SM employing real or complex encoding for a symmetric interference channel model. \\
The techniques considered in this paper are optimal in different operating regimes. We show that the sum outage capacity can be maximized by reducing the SM rate to a value less than the maximum allowed value. The optimum SM rate depends on the number of interferers and the number of available active users. In particular, we show that the generalized multi-user SM (MU SM) method employing real-valued encoding provides a performance that is either comparable, or significantly higher than that of MU SM employing complex encoding. A combination of analysis and simulation is used to describe the trade-off between the multiplexing rate and sum outage capacity for different antenna configurations.
\end{abstract}
\def\bGIQkm{\mathbf{\tilde{G}}^{m}_k}
\def\bC{\mathbf{C}}
\def\bV{\mathbf{V}}
\def\bh{\mathbf{h}}
\def\bw{\mathbf{w}}
\def\bW{\mathbf{W}}
\def\bo{\mathbf{1}}
\def\bs{\mathbf{s}}
\def\bv{\mathbf{v}}
\def\bg{\mathbf{g}}
\def\bG{\mathbf{G}}
\def\bs{\mathbf{s}}
\def\bc{\mathbf{c}}
\def\bn{\mathbf{n}}
\def\bU{\mathbf{U}}
\def\bl{\mathbf{l}}
\def\bk{\mathbf{k}}
\def\bthe{\mathbf{\theta}}
\def\bz{\mathbf{z}}
\def\by{\mathbf{y}}
\def\be{\mathbf{e}}
\def\bx{\mathbf{x}}
\def\bPk{\mathbf{P}_k}
\def\bPf{\mathbf{P} (f)}
\def\bQf{\mathbf{Q} (f)}
\def\bQof{\mathbf{Q_0} (f)}
\def\bDf{\mathbf{D} (f)}
\def\bDof{\mathbf{D_0} (f)}
\def\bDbf{\mathbf{\bar{D}} (f)}
\def\bPcf{\mathbf{P}^{\dagger} (f)}
\def\bh{\mathbf h}
\def\bp{\mathbf p}
\def\bH{\mathbf H}
\def\bP{\mathbf P}
\def\bX{\mathbf X}
\def\bY{\mathbf Y}
\def\bYt{\mathbf{Y}_t}
\def\bT{\mathbf T}
\def\bff{\mathbf f}
\def\bvareps{\bf \varepsilon}
\def\balpha{{\mathbf \alpha}}
\def\bA{\mathbf A}
\def\bbA{\mathbf A}
\def\bB{\mathbf B}
\def\bI{\mathbf I}
\def\bG{\mathbf G}
\def\bQ{\mathbf Q}
\def\bR{\mathbf R}
\def\byk{\mathbf{y}(k)}
\def\bhk{\mathbf{h}(k)}
\def\bgk{\mathbf{g}_{l}(k)}
\def\bw{\mathbf w}
\def\bwk{\mathbf{w}(k)}
\def\bzk{\mathbf{z}(k)}
\def\bnk{\mathbf{n}(k)}
\def\bxk{x(k)}
\def\bxhk{\hat{x}(k)}
\def\bek{\mathbf{e}(k)}
\def\be{\mathbf e}
\def\beIQ{\mathbf{\tilde{e}}}
\def\bbk{b(k)}
\def\bbbk{{\bar{b}}(k)}
\def\bGk{\mathbf{G}(k)}

\def\byIQk{\mathbf{\tilde{y}}(k)}
\def\byIk{\mathbf{\tilde{y}}_{I}(k)}
\def\byQk{\mathbf{\tilde{y}}_{Q}(k)}
\def\bzIQk{\mathbf{\tilde{z}}(k)}
\def\bzIQlek{\mathbf{\tilde{z}}_{\texttt{LE}}(k)}
\def\bhIQk{\mathbf{\tilde{h}}(k)}
\def\bgIQk{\mathbf{\tilde{g}}^{l}(k)}
\def\bwIQk{\mathbf{\tilde{w}}(k)}
\def\bwIk{\mathbf{\tilde{w}}_{I}(k)}
\def\bwQk{\mathbf{\tilde{w}}_{Q}(k)}
\def\bWIQk{\mathbf{\tilde{W}}(k)}
\def\bWIQdk{\mathbf{\tilde{W}}(k)}
\def\bWIQdl{\mathbf{\tilde{W}}(l)}
\def\bWIQlek{\mathbf{\tilde{W}_{\texttt{LE}}}(k)}
\def\bnIQk{\mathbf{\tilde{n}}(k)}
\def\bnIQck{\mathbf{\tilde{n}^{'}}({k+m})}
\def\bsIQk{\mathbf{\tilde{s}}(k)}
\def\bsIQck{\mathbf{\tilde{s}^{'}}({k+m})}
\def\bxIQk{\mathbf{\tilde{x}}(k)}
\def\bxIQhk{\mathbf{\hat{x}}(k)}
\def\beIQk{\mathbf{\tilde{e}}(k)}
\def\beIQck{\mathbf{\tilde{e}^{\dagger}}(k)}
\def\bbIQk{{\tilde{b}}(k)}
\def\bBIQk{\mathbf{\tilde{B}}(k)}
\def\bBIQbk{\mathbf{\bar{B}}(k)}
\def\beIQ{\mathbf{\tilde{e}}}
\def\bHIQ{\mathbf{\tilde{H}}}
\def\bhIQ{{\mathbf{ \tilde{h}}}}
\def\bHIQk{\mathbf{\tilde{H}}(k)}
\def\bhIQk{\mathbf{\tilde{h}}(k)}

\def\bhIQz{\mathbf{\tilde{h}} (z)}
\def\bhIQdz{\mathbf{\tilde{h}^{\dagger}} (z)}
\def\bhIQcz{\mathbf{\tilde{h}^{\dagger}} (z^{-1})}

\def\bhz{\mathbf{h} (z)}
\def\bhdz{\mathbf{h^{\dagger}} (z)}
\def\bhcz{\mathbf{h^{\dagger}} (z^{-1})}

\def\byIQ{\mathbf{\tilde{y}}}
\def\bwIQ{\mathbf{\tilde{w}}}
\def\bWIQ{\mathbf{\tilde{W}}}
\def\bnIQ{\mathbf{\tilde{n}}}
\def\bYIQf{\mathbf{\tilde{Y}} (f)}
\def\bYIQcf{\mathbf{\tilde{Y}^{\dagger}} (f)}
\def\byIQf{\mathbf{\tilde{y}} (f)}
\def\byIQcf{\mathbf{\tilde{y}^{\dagger}} (f)}
\def\bHIQf{\mathbf{\tilde{H}} (f)}
\def\bHIQcf{\mathbf{\tilde{H}^{\dagger}} (f)}
\def\bhIQf{\mathbf{\tilde{h}} (f)}
\def\bomegaIQf{\mathbf{\tilde{\omega}} (f)}
\def\bhIQdf{\mathbf{\tilde{h}^{\dagger}} (f)}
\def\bhIQcf{\mathbf{\tilde{h}^{\dagger}} (f)}
\def\bomegaIQcf{\mathbf{\tilde{\omega}^{\dagger}} (f)}
\def\bhIcf{\mathbf{h^{\dagger}}_{I} (f)}
\def\bhQcf{\mathbf{h^{\dagger}}_{Q} (f)}
\def\bgIQf{\mathbf{\tilde{g}}_{l}(f)}
\def\bgIQcf{{\mathbf{\tilde{g}}^{\dagger}}_l (f)}
\def\bGIQf{\mathbf{\tilde{G}}^{l}(f)}
\def\bGIQcf{{\mathbf{\tilde{G}}^{\dagger}}_l (f)}
\def\bWIQf{\mathbf{\tilde{W}} (f)}
\def\bwIQf{\mathbf{\tilde{w}} (f)}
\def\bwIf{\mathbf{w}_{I} (f)}
\def\bwQf{\mathbf{w}_{Q} (f)}
\def\bWIQdf{\mathbf{\tilde{W}_{\texttt{DFE}}} (f)}
\def\bWIQle{\mathbf{\tilde{W}_{\texttt{LE}}} (f)}
\def\bwIQle{\mathbf{\tilde{w}_{\texttt{LE}}} (f)}
\def\bNIQf{\mathbf{\tilde{n}} (f)}
\def\bNIQcf{\mathbf{\tilde{n}^{\dagger}} (f)}
\def\bnIQf{\mathbf{\tilde{n}} (f)}
\def\bXIQf{\mathbf{\tilde{X}} (f)}
\def\bEIQf{\mathbf{\tilde{E}} (f)}
\def\bEIQcf{\mathbf{\tilde{E}^{\dagger}} (f)}
\def\beIQf{\mathbf{\tilde{e}} (f)}
\def\beIQcf{\mathbf{\tilde{e}^{\dagger}} (f)}
\def\bEIQdf{\mathbf{\tilde{E}_{\texttt{dfe}}} (f)}
\def\bEIQ{\mathbf{\tilde{E}}}
\def\bBIQf{\mathbf{\tilde{B}}(f)}
\def\bBIQcf{\mathbf{\tilde{B}^{\dagger}}(f)}
\def\bYf{\mathbf {Y} (f)}
\def\byf{\mathbf {y} (f)}
\def\bycf{\mathbf {y^{\dagger}} (f)}
\def\bHf{\mathbf {H} (f)}
\def\bhf{\mathbf {h} (f)}
\def\bgf{\mathbf{g}^{m}(f)}
\def\bgcf{\mathbf{g}^{m\dagger} (f)}
\def\bhcf{\mathbf {h^{\dagger}} (f)}
\def\bWf{\mathbf {W} (f)}
\def\bwf{\mathbf {w} (f)}
\def\bWdf{\mathbf{W_{\texttt{DFE}}} (f)}
\def\bWle{\mathbf{W_{\texttt{LE}}} (f)}
\def\bwdf{\mathbf{w_{\texttt{DFE}}} (f)}
\def\bwle{\mathbf{w_{\texttt{LE}}}(f)}
\def\bbdf{{b_{\texttt{DFE}}}(f)}
\def\bnf{\mathbf{n} (f)}
\def\bEf{\mathbf{E} (f)}
\def\bEcf{\mathbf{E^{\dagger}} (f)}
\def\bxf{x(f)}
\def\bxhf{\hat{x}(f)}
\def\bef{\mathbf{e} (f)}
\def\becf{\mathbf{e^{\dagger}} (f)}
\def\bEdf{\mathbf{E_{\texttt{dfe}}} (f)}
\def\bedf{\mathbf{e_{\texttt{dfe}}} (f)}
\def\bE{\mathbfE}
\def\bBf{\mathbf{B} (f)}
\def\bBcf{\mathbf{B^{\dagger}} (f)}
\def\bbf{\mathbf{b} (f)}
\def\bbcf{\mathbf{b^{\dagger}} (f)}

\def\buf{u (f)}
\def\budf{{u^{\dagger}} (f)}
\def\buif{{u^{-1}} (f)}
\def\bucf{{u^{\dagger}} (f)}

\def\bYIQD{\mathbf{\tilde{Y}} (D)}
\def\bYIQcD{\mathbf{\tilde{Y}^{\dagger}} (D^{-*})}
\def\byIQD{\mathbf{\tilde{y}} (D)}
\def\byIQcD{\mathbf{\tilde{y}^{\dagger}} (D^{-*})}
\def\bGIQD{\mathbf{\tilde{G}^{m}} (D)}
\def\bHIQD{\mathbf{\tilde{H}} (D)}
\def\bHIQcD{\mathbf{\tilde{H}^{\dagger}} (D^{-*})}
\def\bhIQD{\mathbf{\tilde{h}} (D)}
\def\bhID{\mathbf{h}_I (D)}
\def\bhQD{\mathbf{h}_Q (D)}
\def\bhIQdD{\mathbf{\tilde{h}^{\dagger}} (D)}
\def\bhIQcD{\mathbf{\tilde{h}^{\dagger}} (D^{-*})}
\def\bhIcD{\mathbf{h^{\dagger}}_{I} (D^{-*})}
\def\bhQcD{\mathbf{h^{\dagger}}_{Q} (D^{-*})}
\def\bgIQD{\mathbf{\tilde{g}} (D)}
\def\bgIQcD{\mathbf{\tilde{g}^{\dagger}} (D^{-*})}
\def\bWIQD{\mathbf{\tilde{W}} (D)}
\def\bwIQD{\mathbf{\tilde{w}} (D)}
\def\bwID{\mathbf{w}_{I} (D)}
\def\bwQD{\mathbf{w}_{Q} (D)}
\def\bWIQdD{\mathbf{\tilde{W}} (D)}
\def\bWIQle{\mathbf{\tilde{W}_{\texttt{LE}}} (D)}
\def\bwIQle{\mathbf{\tilde{w}_{\texttt{LE}}} (D)}
\def\bnIQcD{\mathbf{\tilde{n}^{\dagger}} (D^{-*})}
\def\bnIQD{\mathbf{\tilde{n}} (D)}
\def\bxIQD{\mathbf{\tilde{x}} (D)}
\def\bxIQcD{\mathbf{\tilde{x}}^{\dagger} (D)}
\def\bEIQD{\mathbf{\tilde{E}} (D)}
\def\bEIQcD{\mathbf{\tilde{E}^{\dagger}} (D^{-*})}
\def\beIQD{\mathbf{\tilde{e}} (D)}
\def\beIQcD{\mathbf{\tilde{e}^{\dagger}} (D^{-*})}
\def\bEIQdD{\mathbf{\tilde{E}_{\texttt{dfe}}} (D)}
\def\bEIQ{\mathbf{\tilde{E}}}
\def\bBIQD{\mathbf{\tilde{B}}(D)}
\def\bBIQcD{\mathbf{\tilde{B}^{\dagger}}(D^{-*})}
\def\bYD{\mathbf {Y} (D)}
\def\byD{\mathbf {y} (D)}
\def\bycD{\mathbf {y^{\dagger}} (D^{-*})}
\def\bHD{\mathbf {H} (D)}
\def\bhD{\mathbf {h} (D)}
\def\bgD{\mathbf {g} (D)}
\def\bhcD{\mathbf {h^{\dagger}} (D^{-*})}
\def\bhIcD{\mathbf {h^{\dagger}}_{I} (D^{-*})}
\def\bhQcD{\mathbf {h^{\dagger}}_{Q} (D^{-*})}
\def\bgcD{\mathbf {g^{\dagger}} (D^{-*})}
\def\bWD{\mathbf {W} (D)}
\def\bwID{\mathbf {w}_{I} (D)}
\def\bwQD{\mathbf {w}_{Q} (D)}
\def\bwD{\mathbf {w} (D)}
\def\bWdD{\mathbf{W_{\texttt{DFE}}} (D)}
\def\bWle{\mathbf{W_{\texttt{LE}}} (D)}
\def\bwdD{\mathbf{w_{\texttt{DFE}}} (D)}
\def\bwle{\mathbf{w_{\texttt{LE}}}(D)}
\def\bbdD{{b_{\texttt{DFE}}}(D)}
\def\bnD{\mathbf{n} (D)}
\def\bncD{\mathbf {n^{\dagger}} (D^{-*})}
\def\bED{\mathbf{E} (D)}
\def\bEcD{\mathbf{E^{\dagger}} (D^{-*})}
\def\bxD{x(D)}
\def\bxhD{\hat{x}(D)}
\def\beD{\mathbf{e} (D)}
\def\becD{\mathbf{e^{\dagger}} (D^{-*})}
\def\bEdD{\mathbf{E_{\texttt{dfe}}} (D)}
\def\bedD{\mathbf{e_{\texttt{dfe}}} (D)}
\def\bE{\mathbfE}
\def\bBD{\mathbf{B} (D)}
\def\bBcD{\mathbf{B^{\dagger}} (D^{-*})}
\def\bbD{b (D)}
\def\bbcD{\mathbf{b^{\dagger}} (D^{-*})}
\def\bGD{\mathbf{G} (D)}
\def\bgD{\mathbf{g} (D)}

\def\buD{u (D)}
\def\budD{{u^{\dagger}} (D^{-*})}
\def\buiD{{u^{-1}} (D)}
\def\bucD{{u^{\dagger}} (D)}

\def\buIQD{{\tilde{u}} (D)}
\def\buIQcD{\tilde{u}^{\dagger} (D)}
\def\buIQiD{\tilde{u}^{-1} (D)}
\def\buIQdD{\tilde{u}^{\dagger} (D^{-*})}
\def\buIQdiD{{\tilde{u}^{\dagger{-1}}}(D^{-*})}

\def\bbIQD{\tilde{b} (D)}
\def\bbIQcD{\tilde{b}^{\dagger} (D^{-*})}
\def\bReeIQ{\mathbf{R_{\tilde{e}\tilde{e}}}}
\def\bRee{\mathbf{R_{ee}}}
\def\bReeIQi{\mathbf{R^{-1}_{\tilde{e}\tilde{e}}}}
\def\bRxxD{R_{xx} (D)}
\def\bRxixD{R_{x^{i}x^{i}} (D)}
\def\bRxixiD{R^{-1}_{x^{i}x^{i}} (D)}
\def\bRxyD{\mathbf{R_{xy}} (D)}
\def\bRyyD{\mathbf{R_{yy}} (D)}
\def\bRyyDi{\mathbf{R^{-1}_{yy}} (D)}
\def\bRnnD{\mathbf{R_{nn}} (D)}
\def\bRiiD{\mathbf{R_{ii}} (D)}
\def\bRinD{\mathbf{R_{(i+n)}} (D)}
\def\bRiniD{\mathbf{R^{-1}_{(i+n)}} (D)}
\def\bRnniD{\mathbf{R^{-1}_{nn}} (D)}
\def\bRxxiD{R^{-1}_{xx} (D)}
\def\bRyyiD{\mathbf{R^{-1}_{yy}} (D)}
\def\bRxxIQD{\mathbf{R_{\tilde{x}\tilde{x}}} (D)}
\def\bRxyIQD{\mathbf{R_{\tilde{x}\tilde{y}}} (D)}
\def\bRyyIQD{\mathbf{R_{\tilde{y}\tilde{y}}} (D)}
\def\bRnnIQD{\mathbf{R_{\tilde{n}\tilde{n}}} (D)}
\def\bRiiIQD{\mathbf{R_{\tilde{i}\tilde{i}}} (D)}
\def\bRiiIQf{\mathbf{R_{\tilde{i}\tilde{i}}} (f)}
\def\bRxxIQiD{\mathbf{R^{-1}_{\tilde{x}\tilde{x}}} (D)}
\def\bRnnIQiD{\mathbf{R^{-1}_{\tilde{n}\tilde{n}}} (D)}
\def\bRyyIQiD{\mathbf{R^{-1}_{\tilde{y}\tilde{y}}} (D)}

\def\bUIQD{\mathbf{\tilde{U}}(D)}
\def\bUIQiD{\mathbf{\tilde{U}^{-1}}(D)}
\def\bUIQcD{\mathbf{\tilde{U}^{\dagger}}(D)}
\def\bUIQdD{\mathbf{\tilde{U}^{\dagger}}(D^{-*})}
\def\bUIQdiD{\mathbf{\tilde{U}^{\dagger{-1}}}(D^{-*})}

\def\buID{\mathbf {u_{I}} (D)}
\def\buQD{\mathbf {u_{Q}} (D)}
\def\buIcD{\mathbf {u^{\dagger}_{I}} (D^{-*})}
\def\buQcD{\mathbf {u^{\dagger}_{Q}} (D^{-*})}
\def\buIQcD{\mathbf {\tilde{u}^{\dagger}} (D^{-*})}
\def\buIQdD{\tilde{u}^{\dagger} (D^{-*})}
\def\buIQdiD{\mathbf{\tilde{u}^{\dagger{-1}}}(D^{-*})}

\def\bUIQf{\mathbf{\tilde{U}}(f)}
\def\bUIQif{\mathbf{\tilde{U}^{-1}}(f)}
\def\bUIQcf{\mathbf{\tilde{U}^{\dagger}}(f)}
\def\bUIQdf{\mathbf{\tilde{U}^{\dagger}}(f)}
\def\bUIQdif{\mathbf{\tilde{U}^{\dagger{-1}}}(f)}

\def\buIQf{{\tilde{u}} (f)}
\def\buIQcf{\tilde{u}^{\dagger} (f)}
\def\buIQif{\tilde{u}^{-1} (f)}
\def\buIQdf{\tilde{u}^{\dagger} (f)}
\def\buIQdif{{\tilde{u}^{\dagger{-1}}}(f)}

\def\buIQz{{\tilde{u}} (z)}
\def\buIQcz{\tilde{u}^{\dagger} (z)}
\def\buIQiz{\tilde{u}^{-1} (z)}
\def\buIQdz{\tilde{u}^{\dagger} (z^{-1})}
\def\buIQdiz{{\tilde{u}^{\dagger{-1}}}(z^{-1})}

\def\bbIQf{\tilde{b} (f)}
\def\bbIQcf{\tilde{b}^{\dagger} (f)}
\def\bReeIQ{\mathbf{R_{\tilde{e}\tilde{e}}}}
\def\bRee{\mathbf{R_{ee}}}
\def\bReeIQi{\mathbf{R^{-1}_{\tilde{e}\tilde{e}}}}
\def\bReeIQl{\mathbf{\bar{R}_{\tilde{e}\tilde{e}}}}
\def\bReeIQli{\mathbf{\bar{R}^{-1}_{\tilde{e}\tilde{e}}}}
\def\bRxx{R_{xx} (f)}
\def\bRxix{R_{x^{i}x^{i}} (f)}
\def\bRxixi{R^{-1}_{x^{i}x^{i}} (f)}
\def\bRxy{\mathbf{R_{xy}} (f)}
\def\bRyy{\mathbf{R_{yy}} (f)}
\def\bRnn{\mathbf{R_{nn}} (f)}
\def\bRii{\mathbf{R_{ii}} (f)}
\def\bRin{\mathbf{R_{(i+n)}} (f)}
\def\bRini{\mathbf{R^{-1}_{(i+n)}} (f)}
\def\bRnni{\mathbf{R^{-1}_{nn}} (f)}
\def\bRni{\mathbf{R^{-1}_{nn}}}
\def\bRnib{\mathbf{\bar{R}^{-1}_{nn}}}
\def\bRn{\mathbf{R_{nn}}}
\def\bRnb{\mathbf{\bar{R}_{nn}}}
\def\bRxxi{R^{-1}_{xx} (f)}
\def\bRyyi{\mathbf{R^{-1}_{yy}} (f)}
\def\bRxxIQ{\mathbf{R_{\tilde{x}\tilde{x}}} (f)}
\def\bRxyIQ{\mathbf{R_{\tilde{x}\tilde{y}}} (f)}
\def\bRyyIQ{\mathbf{R_{\tilde{y}\tilde{y}}} (f)}
\def\bRnnIQ{\mathbf{R_{\tilde{n}\tilde{n}}} (f)}
\def\bRnnIQb{\mathbf{\bar{R}_{\tilde{n}\tilde{n}}} (f)}
\def\bRiiIQ{\mathbf{R_{\tilde{i}\tilde{i}}} (f)}
\def\bRxxIQi{\mathbf{R^{-1}_{\tilde{x}\tilde{x}}} (f)}
\def\bRnnIQi{\mathbf{R^{-1}_{\tilde{n}\tilde{n}}} (f)}
\def\bRnnIQib{\mathbf{\bar{R}_{\tilde{n}\tilde{n}}} (f)}
\def\bRyyIQi{\mathbf{R^{-1}_{\tilde{y}\tilde{y}}} (f)}

\def\byIQbfk{\mathbf{\bar{y}} (f_k)}
\def\byIQfk{\mathbf{\tilde{y}} (f_k)}
\def\bhIQfk{\mathbf{\tilde{h}} (f_k)}
\def\bgIQfk{\mathbf{\tilde{g}} (f_k)}
\def\bnIQfk{\mathbf{\tilde{n}} (f_k)}
\def\bhIQcfk{\mathbf{\tilde{h}^{\dagger}} (f_k)}
\def\bgIQcfk{\mathbf{\tilde{g}^{\dagger}} (f_k)}
\def\bnIQcfk{\mathbf{\tilde{n}^{\dagger}} (f_k)}
\def\bRnnIQik{\mathbf{R^{-1}_{\tilde{n}\tilde{n}}} (f_k)}

\def\bUfm{\mathbf{U} (f_m)}
\def\bUcfm{\mathbf{U^{\dagger}} (f_m)}
\def\blamdafm{\mathbf{\Lambda} (f_m)}
\def\bhbIQfm{\mathbf{\bar{h}} (f_m)}
\def\bhbIQcfm{\mathbf{\bar{h}^{\dagger}} (f_m)}

\def\bhIQfm{\mathbf{\tilde{h}} (f_m)}
\def\bnIQfm{\mathbf{\tilde{n}} (f_m)}
\def\bhIQcfm{\mathbf{\tilde{h}^{\dagger}} (f_m)}
\def\bnIQcfm{\mathbf{\tilde{n}^{\dagger}} (f_m)}
\def\bRnnIQim{\mathbf{R^{-1}_{\tilde{n}\tilde{n}}} (f_m)}

\def\bh{{\mathbf h}}
\def\bE{{\mathbf E}}
\def\bg{{\mathbf g}}
\def\bomega{{\mathbf \omega}}
\def\bU{{\mathbf U}}
\def\bp{{\mathbf p}}
\def\bff{{\mathbf f}}
\def\bvareps{{\mathbf \varepsilon}}
\def\balpha{{\mathbf \alpha}}
\def\bA{{\mathbf A}}
\def\bB{{\mathbf B}}

\def\bh{{\mathbf h}}
\def\bp{{\mathbf p}}
\def\bH{{\mathbf H}}
\def\bP{{\mathbf P}}
\def\PIQ{{\tilde{ P}}}
\def\bff{{\mathbf f}}
\def\bvareps{\bf \varepsilon}
\def\balpha{{\mathbf \alpha}}
\def\bA{{\mathbf A}}
\def\bB{{\mathbf B}}
\def\bI{{\mathbf I}}
\def\bQ{{\mathbf Q}}

\def\bh{{\mathbf h}}
\def\bp{{\mathbf p}}
\def\bH{{\mathbf H}}
\def\bP{{\mathbf P}}
\def\bff{{\mathbf f}}
\def\bvareps{\bf \varepsilon}
\def\balpha{{\mathbf \alpha}}
\def\bA{{\mathbf A}}
\def\bB{{\mathbf B}}
\def\bI{{\mathbf I}}
\def\bQ{{\mathbf Q}}

\def\bSf{\mathbf {S} (f)}
\def\bLf{\mathbf {L} (f)}
\def\bLcf{\mathbf {L^{\dagger}} (f)}
\def\bD{\mathbf {D} }
\def\bSz{\mathbf {S} (z)}
\def\bLz{\mathbf {L} (z)}
\def\bLiz{\mathbf {L^{-1}} (z)}
\def\bLcz{\mathbf {L^{\dagger}} (z^{-1})}
\def\bL{\mathbf {L} }
\def\bZ{\mathbf {Z} }
\def\bP{\mathbf {P} }
\def\bE{\mathbf {E} }
\def\bKp{\mathbf {Kp} }
\def\bX{\mathbf {X} }
\def\bF{\mathbf {F} }
\def\bT{\mathbf {T} }
\def\bN{\mathbf {N} }
\def\bR{\mathbf {R} }
\def\bK{\mathbf {K} }
\def\bLi{\mathbf {Li} }
\def\bFt{\mathbf {Ft} }


\section{Introduction}
Interference alignment (IA) \cite{Cadambe:2008}-\nocite{Maddah:2008}\nocite{Jafar:2009} \nocite{Slock:IA}\cite{Motahari:2009} techniques have been proposed as a means to achieve the optimum degrees-of-freedom (DOF) of an interference channel. This technique relies on symbol extension over multiple time/frequency epochs, together with channel state feedback for precoding. Even for the simple case of $K=3$ transmitters, optimal DOF can only be attained by expanding the symbol set over infinitely large number of time/frequency epochs. Approaching DOF with limited symbol set still remains an open problem.\\

Non-circular signals play an important role in increasing the capacity of interference channels. Recently, \cite{Cadambe:2010} has shown that non-circular signals offer higher DOF compared to conventionally used circularly symmetric signals. Based on this result, \cite{Inkyu:12} proposed asymmetric complex-valued signaling together with multi-user diversity (MUD) \cite{Knopp:ICC95}, \cite{Tse:2002} to obtain 1.5 DOF for $K=3$ transmitter case. This method relies on transmitter precoding and MUD to obtain the required gains. In \cite{Slock:IA}, conditions for satisfying IA solutions are derived for K-user MIMO interference channel. This method requires global channel state information at the transmitter for enabling interference mitigation at the receiver. Additionally, \cite{Ho} considered non-circular complex Gaussian signaling with two users where real-valued signaling is shown to provide optimum sum rate. \\

In \cite{Jung:12}, an opportunistic interference nulling (OIN) method is proposed for the uplink with $K$ base stations (BSs) each with $M$ antennas, and $M$ single antenna users simultaneously communicating with their own BSs. In this method, each BS opportunistically selects a set of users who generate the minimum interference to the other BSs. It is shown that $KM$ DOFs are achievable under the OIN protocol, if the total number of active users in a BS scales at least as $SNR^{(K-1)M}$ where $SNR$ is the operating signal-to-noise power ratio. This work is further generalized in \cite{Paul:IA} for the case of user having $N$ antennas. It is shown that a singular value decomposition (SVD)-based OIN method can reduce the required users to $SNR^{(K-1)M-N+1}$ by optimizing weight vectors at each user.\\

In this paper, we consider $K$ transmitter downlink interference channel in which all the transmitters employ spatial multiplexing (SM) \cite{Telatar:June95}-\nocite{Foschini:Mar98}\nocite{Greedy:Oct05}\nocite{Peel:Jan05}\cite{Peel:Mar05} using $N_t$ antennas. Each transmitter simultaneously serves a group of $N_t$ users that are selected from a pool of $L$ active users. We consider two transmission formats employing either complex or real encoding. Performance analysis is carried out for each case independently. Every user in the system is assumed to have $N_r$ receiver antennas. Our analysis is general and encompasses the special case of single antenna at the transmitter and receiver. In the considered framework, every active user periodically reports the post-processing signal-to-interference-plus-noise-power-ratio (post-SINR) of the receiver to the serving transmitter. We consider a particular approach where interference is mitigated in two stages. In the first stage, the receiver exploits multiple receiver antennas to suppress a portion of the interference. In the second stage, an opportunistic scheduler selects a group of users with highest sum rate so that the scheduled users become nearly free of interference. We provide analytical results quantifying the interference suppression gain and outage capacity using a successive max-SINR scheduler. To this end, we consider a symmetric interference channel (SIC) model where the power level of the dominant interferers are assumed to be equal while the weak interference is treated as additive white Gaussian noise (AWGN); this assumption is also used in \cite{Jung:12}, \cite{Paul:IA} to study DOF in interference limited networks.\\

We analyze the performance for the following operational scenarios: a) Single stream transmission (SST) with complex-valued encoding  b) MU SM with complex-valued encoding c) SST with real-value encoding and its generalization to MU SM employing real-valued encoding. The analysis is first carried out for the SST modes employing either complex or real encoding and these results are further generalized to MU SM cases.\\

In section \ref{sec:2}, first we analyze the performance of SST technique with max-SINR scheduling. We show that opportunistic scheduling exploits \emph{spatial interference alignment} (SIA) property inherent to a multi-user system for effective interference mitigation. More specifically, we show that the post-SINR of the scheduled user employing multiple antenna minimum-means-square-estimation (MMSE) receiver reaches a high value when the interference covariance matrix (ICM) of the scheduled user becomes nearly rank deficient. This condition generally occurs when the scheduler selects a particular user whose interfering channel vectors tend to become linearly dependent. This phenomenon is referred to as SIA. We obtain a tight bound for the outage probability and the sum outage capacity which shows that we get a sum outage capacity of $K \log(1+SNR)$ bits/sec/Hz when the number of active users $L$ is proportional to $SNR^{K-N_r}$. In section \ref{sec:3}, the analysis for SST with complex-encoding is further generalized to the general case of MU SM employing complex encoding. Using a sub-optimal successive max-SINR scheduling algorithm, we show that a sum outage capacity of $KN_t \log(1+SNR)$ can be obtained when $L \propto SNR^{KN_t-N_r}$.\\

In section \ref{sec:4}, we propose a transmission method which is suitable for systems with limited antennas (including the single antenna case) and a large number of interferers. In the proposed system model, all the transmitters in the network transmit \emph{a single data stream} using \emph{real-valued} modulation alphabets. The receiver at each user collects the real and imaginary parts of the multi-antenna receiver to obtain a virtual antenna array of size $2N_r$. The receiver further filters the real and imaginary parts of the received signal using a widely linear (WL) MMSE filter \cite{picinbono:95}-\nocite{Gerstacker:Sept03} \nocite{Chevaliar:Mar2006}\cite{Kuchi:Jan09}  for data detection. We show that this method offers a sum outage capacity of $\frac{K}{2} \log(1+SNR)$ when $L \propto SNR^{\frac{K}{2}-N_r}$. This result is further generalized to MU SM employing real-encoding. We show that by spatial multiplexing $t$ real-valued data streams using $t$ antennas, we  get a sum outage capacity of $\frac{tK}{2} \log(1+SNR)$ when $L \propto SNR^{\frac{tK}{2}-N_r}$. While complex-valued MU SM offers a SM rate of $R=N_t$ where $N_t$ takes integer values, real-valued MU SM gives fractional multiplexing rates of $R=\frac{t}{2}$ which take values in steps of $0.5$. The real-valued encoder can be viewed as a generalized SM encoder. Using $t=2N_t$, we get the same user scaling results as that of complex-valued encoding. Numerical results are given in section \ref{sec:5} where we illustrate the trade-off between the SM rate and achievable capacity. Finally, conclusions are drawn in section \ref{sec:7}.

\section{System Model for Single Stream Transmission}\label{sec:2}
We consider $K$ single antenna transmitters each with $L$ single antenna active users. All transmitters simultaneously send a single complex-valued data stream to one of the $L$ users. The baseband received signal for the user with index $l$ that is served by a given transmitter is represented as
\begin{eqnarray}
\by_{l}(k) =\sqrt{S}  \bh_{l} x_{l}(k) + \sum_{i=1}^{K-1} \sqrt{I_0} \bg_{i,l} x_{i,l}(k) +\bn_{l}(k), \quad l=1,2,..,L
\end{eqnarray}
where $k$ denotes discrete time index, $S$ is the signal power, and $I_0$ denotes the power level of each individual interferer. The desired and interfering signal channel vectors $\bh_{l}$ and $\bg_{i,l}$ for each $i$ are modeled as multivariate circularly symmetric complex Gaussian
random vectors having independent, identically distributed (i.i.d.) elements with $\textsf{E}[\bh_{l}]=\textsf{E}[ \bg_{i,l}]=\mathbf{0}$ and $\textsf{E}[\bh_{l}
\bh^{\dagger}_{l}]=\textsf{E}[ \bg_{i,l}\bg^{\dagger}_{i,l}]=\bI$, where $\textsf{E}$ denotes the expectation operation and $\bI$ denotes the identity matrix. The noise term $\bn_{l}$ is modeled as a circularly
symmetric complex Gaussian noise vector composed of i.i.d. elements with zero-mean and variance $\frac{N_0}{2}$ per dimension. The operating signal-to-noise power ratio (SNR) is defined as: $SNR=\frac{S}{N_0}$. The complex-valued modulation sequences $x_{l}(k)$ and $x_{i,l}(k)$ are assumed to be  i.i.d. circularly symmetric complex Gaussian random variables (r.v's) with zero-mean, unit variance, and statistically independent of each other.
\subsection{Max-SINR Scheduling based on Post-processing SINR of MMSE}
The MMSE receiver weighs and combines the $N_r$ copies of the received signal samples using an un-biased MMSE filter \cite{Cioffi:Stanford} $\bw_{l}$ to produce a decision variable $z_{l}(k)=\bw_{l}
\by_{l}(k)$, where $\bw_{l}=\sqrt{S}\bh^{\dagger}_{l} \hat{\bR}^{-1}_{l}$
and
\begin{eqnarray*}
\hat{\bR}_{l}= \textsf{E} \left(\sum_{i=1}^{K-1} \sqrt{I_0} \bg_{i,l} x_{i,l}(k) + \bn_{l}(k)\right) \left(\sum_{i=1}^{K-1} \sqrt{I_0} \bg_{i,l} x_{i,l}(k) + \bn_{l}(k)\right )^{\dagger}=\sum_{i=1}^{K-1} I_0 \bg_{i,l} \bg^{\dagger}_{i,l} + N_{0}\bI
\end{eqnarray*}
denotes the short-term noise-plus interference covariance matrix (NICM). The SINR at the output of the MMSE receiver is given by:
$
\gamma_{l} =  S \bh^{\dagger}_{l} \hat{\bR}^{-1}_{l} \bh_{l}.
$ Let $\bR_{l} =  \sum_{i=1}^{K-1} I_0  \bg_{i,l} \bg^{\dagger}_{i,l}$ be the interference covariance matrix (ICM) and the symbol $\dagger$ denotes conjugate-transpose operation.\\
In the proposed scheduling policy, each transmitter allocates the entire available bandwidth to the user with highest reported instantaneous post-SINR. The transmitter serves the user with index $l^{*}$ with maximum reported
post-SINR i.e.,$ \gamma_{l^{*}} = \texttt{max} \left(\gamma_{1}, \gamma_{2},..,\gamma_{L} \right) $. The transmitter selects a suitable modulation and coding technique and
transmits to the scheduled user at a rate $I=\log(1+\gamma_{l^{*}})$\footnote{Use of capacity achieving codes with large block lengths is assumed here.} where the logarithm is taken with respect to base 2. We introduce a metric called \emph{transmitter outage probability} (TOP) which is defined as: $ P_{\texttt{out}}= P(I < \log(1+\beta)) $ where $\log(1+\beta)$ is the target \texttt{outage capacity} of the transmitter and $\beta$ is a \texttt{target SNR} which is distinct from the operating SNR.\\
Further, the TOP can be expressed in alternative form as:
$
P_{\texttt{out}} = P( \texttt{max} \left(\gamma_{1}, \gamma_{2},..,\gamma_{L} \right) < \beta)
$. To simplify the analysis, we express the post-SINR in an alternative form. Let $r_{l}$ denote the rank of ICM. Then $\bR_{l}$ has exactly $r_{l}$ positive ordered eigenvalues represented in vector form:
$\lambda_{l}=[\lambda_{l,1},\lambda_{l,2},..,\lambda_{l,r_{l}}]$   ($\lambda_{l,1}>\lambda_{l,2}>..>\lambda_{l,r_{l}}$), and the remaining $N_r-r_{l}$ eigenvalues are identically equal to zero. Therefore, the eigenvalues of $\hat{\bR}_{l}$ can
be expressed as: $ \hat{\lambda}_{l,p} = \lambda_{l,p} + N_{0},\,\,$ for $\,\, p=1,..,r_{l}$ and $ \hat{\lambda}_{l,p}= N_{0}, \,\,$ for $\,\, p=r_{l}+1,..,N_r$. The rank of ICM can be expressed as: $r_{l}=\texttt{min}(N_r,K-1)$. Next, the matrix $\hat{\bR}_{l}$ is expressed as:
$
\hat{\bR}_{l}=\bU^{\dagger}_{l} \hat{\Lambda}^{-1}_{l} {\bU}_{l}
$
where $\hat{\Lambda}_{l}=\texttt{diag}\left[\hat{\lambda}_{l,1},\hat{\lambda}_{l,2},..,\hat{\lambda}_{l,N_r}\right]$ is a diagonal matrix of size $N_r \times N_r$ and $\bU_{l}$ represents a unitary matrix. Using this, the post-SINR can be
expressed as
\begin{eqnarray}
\gamma_{l}  = S  {\omega^{\dagger}}_{l}  \hat{\Lambda}^{-1}_{l} \omega_{l} =   S \sum_{p=1}^{r_{l}} \frac{|\omega_{l,p}|^2}{\lambda_{l,p}+N_{0}}+S\sum_{p=r_{l}+1}^{N_r} \frac{|\omega_{l,p}|^2}{N_{0}} , \quad l=1,2,..,L\label{Eq:postEigen}
\end{eqnarray}
where $\omega_{l}=[\omega_{l,1}, \omega_{l,2},..,\omega_{l,N_r}]^{T}=\bU_{l} \bh_{l}$.\\
When the number of interferers ($K-1$) is less than $N_r$, the ICM becomes rank deficient. As $N_0\rightarrow0$, the post-SINR
scales inversely with $N_0$ for all users in the system. In the high SNR limit, the MMSE receiver suppresses all $K-1$ interferers as long as $K-1< N_r$. In the opposite case, for $K-1 \ge N_r$, the ICM has full rank. In this case the post-SINR is dictated by the instantaneous eigenvalues of the ICM. Though MMSE receiver by itself cannot provide full interference suppression all the time, it leads to additional interference suppression gain when an
opportunistic scheduler selectively schedules a user with maximum instantaneous post-SINR. In the following, we consider performance analysis for the case when $K-1 \ge N_r$ which is of interest to us. In the case of $K-1 <N_r$, the MMSE receiver provides full interference suppression leading to noise limited case, while the max-SINR scheduler provides further multi-user diversity gain. In \cite{Tse:2002}, it has been shown that the capacity increases as $\ln(L)\log(1+SNR)$.\\

For the case of $K-1 \ge N_r$, the post-SINR can be expressed as: $ \gamma_{l} =  S \sum_{p=1}^{N_r}
\frac{|\omega_{l,p}|^2}{\lambda_{l,p}+N_{0}}. $ The summation is typically dominated by the last term corresponding to the minimum eigenvalue i.e.,
\begin{eqnarray}
\gamma_{l} &\ge & S \frac{|\omega_{l,N_r}|^2}{\lambda_{l,N_r}+N_{0}}. \label{Eq:LBsnr}
\end{eqnarray}
For most channel realizations, the ICM generally has full rank i.e., the interfering channel vectors are linearly independent for most users. However, the ICM becomes rank deficient i.e., $\lambda_{l,N_r}=0$  when  a subset of interference channel vectors  (ICV) align to a common direction or, more generally the interference channel vectors (ICV) become linearly dependent. We refer to this phenomenon as \emph{spatial interference alignment} (SIA). When the number of active users is very high, each transmitter is likely to encounter a few users that have rank deficient ICM. Since the channels take values from a continuous probability distribution, the probability of a small set of discrete events wherein a subset of ICV becoming linearly dependent is zero. However, in practice, it is not necessary to meet the rank deficient criterion strictly to achieve high capacity. What is more important is that the minimum eigenvalue takes a value smaller than noise power level i.e., $\lambda_{l,N_r} < N_{0}$. Alternatively, if $\lambda_{l,N_r}=\epsilon$ where $\epsilon <  N_{0}$ and $N_0 \rightarrow 0$, we term this condition as \emph{$\epsilon$ spatial interference alignment} or simply \emph{spatial interference alignment}.\\

\subsubsection*{Connections to Explicit Interference Alignment}
Ref [1] uses symbol extension and applies a set of weights on the repeated symbols such that the interference channel vectors are aligned at the receiver. A ZF or MMSE receiver exploits the IA property for signal separation. Explicit IA requires the users to feedback exact value of signal and interference channel vectors. However, the framework proposed here requires significantly reduced feedback in the form of post-SINR. Our approach relies on the fact that one user selected from a large pool obeys $\epsilon$ SIA with high probability. Essentially, we rely on multi-user diversity to provide required interference mitigation.\\

Next, we obtain a closed-form expression for the TOP using the lower bound (\ref{Eq:LBsnr}) on post-SINR, which leads to an upper-bound (UB) on the TOP. If we assume that each user reports the lower bound (LB) on SINR given by (\ref{Eq:LBsnr}) instead of the actual SINR, the TOP is upper-bounded as:
$
P_{\texttt{out},\texttt{UB}} = P( \texttt{max} \left(\bar{\gamma}_{1}, \bar{\gamma}_{2},..,\bar{\gamma}_{L} \right) < \beta)
$
where $\bar{\gamma}_{l}=S  \frac{|\omega_{l,N_r}|^2}{\lambda_{l,N_r}+N_{0}}$. Since $\bar{\gamma}_{l}$ are i.i.d. r.v's, the TOP-UB is expressed as
\begin{eqnarray}
P_{\texttt{out},\texttt{UB}} = F^{L}(\beta) \label{Eq:ConvOutage}
\end{eqnarray}
where $F(\beta)=P(\bar{\gamma}_{l}< \beta)$ and,
\begin{eqnarray}
F(\beta)= P\left(S  \frac{x}{\lambda_{m}+N_{0}} < \beta \right) \label{Eq:outagelb}
\end{eqnarray}
where $\lambda_{m}=\lambda_{l,N_r}$ denotes the minimum eigenvalue (MEV) and we omit the dependency on $l$ for notational simplicity. Let $x=|\omega_{l,N_r}|^2$ be the $N_r$th element of the vector $\omega_{l}$. The projection of a zero-mean i.i.d. complex Gaussian random vector on to an unitary matrix
gives another r.v. with zero-mean i.i.d. complex Gaussian distribution. Therefore, the vector $\omega_{l}=\bU_{l} \bh_{l}$ has same distribution as $\bh_{l}$. As a consequence, the r.v. $x$ has exponential distribution: $ p(x)= e^{-x}, \quad x \ge 0 $ with cumulative distribution function (cdf): $ p(x < a)= 1-e^{-a}, \quad a \ge 0. $\\
Recall that the ICM is defined as: $\bR_{l}=\sum_{i=1}^{K-1} I_0  \bg_{i,l} \bg^{\dagger}_{i,l}$ where  $\bg_{i,l} $ is a zero-mean i.i.d. complex Gaussian vector with
covariance $\bI$. This matrix is called complex central Wishart matrix \cite{Edelman}, \cite{Chiani:Feb03} and its distribution is denoted by ${\emph{CW}}_{m}(n, I_0 \bI)$), $n\ge m$ where $m=\texttt{min}(N_r,K-1)$
and $n=\texttt{max}(N_r, K-1)$. Since we are considering the specific case of $K-1 \ge N_r$, we have: $m=N_r$, $n=K-1$. The joint probability density function (pdf) of ordered eigenvalues
$\lambda_1>\lambda_2>..>\lambda_m>0$ is given in \cite{Wishart:Natraj}\nocite{Kwan:07}\nocite{Chinai:April09}-\cite{Chinai:12}. Using this, the pdf of minimum eigenvalue can be evaluated. The evaluation is straightforward for small values of $N_r,K-1$ and it gets tedious for large values. However, closed-form expression for this pdf is available in polynomial form as \cite{Edelman}, \cite{book:Murihead}, \cite{Khatri:64}
\begin{eqnarray}
p(\lambda_m)=  \frac{1}{I_0} e^{\frac{-m\lambda_m}{I_0}} \sum_{k=k_0}^{K_0} a(k) \left(\frac{\lambda_m}{I_0}\right)^k \label{Eq:pdfmin}
\end{eqnarray}
where elements of $a(k)$ can be obtained using either the mathematica program given in the Appendix of \cite{Edelman} or the closed-form expression given in \cite{Wishart:Natraj}. Table I provides the values for certain combinations of $(n,m)$. Here, $k_0=K-N_r-1$ and $K_0$ is equal to the number of non-zero elements of the vector $a(k)$.  \\
For the special case of $N_r=K-1$ where the number of interferers is equal to the antenna array size, $\lambda_m$ has exponential distribution
\begin{eqnarray*}
p(\lambda_m)= \frac{N_r}{I_0}e^{\frac{-N_r \lambda_m}{I_0}}
\end{eqnarray*}
with mean $\textsf{E} (\lambda_m)=\frac{I_0}{N_r}$. In this case, the pdf has its peak at $\lambda_m=0$ which implies that $P(0 < \lambda_m <N_0)$ is significantly high for small values of $N_0$. It also implies that when the number of active users $L$ is sufficiently large, the scheduled user with $N_r$ receiver antennas can fully reject $N_r$ interferers. The exact number of users required to meet this condition depends on the $\epsilon$ SIA probability $P(0< \lambda_m <N_0)$.  Note that this probability decreases quickly for $K-1 > N_r$, since the pdf expression given in (\ref{Eq:pdfmin}) vanishes at $\lambda_m=0$ for $K-1 > N_r$. Therefore, the number of active users required to fulfill this condition will be very large when $K-1-N_r$ takes high values. The exact number of users required for achieving full interference suppression is determined by evaluating the TOP in
closed-form. We carry out this exercise for the general case involving $N_r$ antennas and $K-1$ interferers.
\subsubsection{TOP Evaluation}\label{TOPcomplex}
First, we begin with (\ref{Eq:outagelb})
\begin{eqnarray}
F(\beta)=\int_{\lambda_m=0}^{\infty} P \left(x < \beta \frac{(\lambda_{m}+N_{0})}{S}\right)
p(\lambda_m) \quad  d \lambda_m = 1- e^{\frac{-\beta N_{0}}{S}} \sum_{k=k_0}^{K_0} \frac{a(k) k!}{\left(\frac{\beta I_0}{S}+m
\right)^{k+1}}.
\end{eqnarray}
The UB on TOP is:
$
P_{\texttt{out},\texttt{UB}} = \left[ 1- e^{\frac{-\beta N_{0}}{S}} \sum_{k=k_0}^{K_0}  \frac{a(k) k!}{\left(\frac{\beta I_0}{S}+m
\right)^{k+1}} \right ]^{L}.
$
Alternatively, the number of active users required to meet a given TOP is given by
\begin{eqnarray}
L=\frac{\ln (P_{\texttt{out},\texttt{UB}})}{\ln \left[ 1- e^{\frac{-\beta N_{0}}{S}} \sum_{k=k_0}^{K_0}  \frac{a(k) k!}{\left(\frac{\beta I_0}{S}+m \right)^{k+1}}
\right ]}. \label{Eq:Lexact}
\end{eqnarray}
For large values of  $\frac{\beta I_0}{S}+m$,  $L$ is approximated as:
$
L \approx e^{\frac{\beta N_{0}}{S}} \ln (P^{-1}_{\texttt{out},\texttt{UB}}) \frac{\left(\frac{\beta I_0}{S}+ m\right)^{K-N_r}}{a(k_0) (k_0)!}
$
where we use the approximation: $\ln(1-x) \approx -x$ and retain only the first term in the summation. When $\frac{\beta I_0}{S}+m$ is large and when $\frac{\beta I_0}{S} \ge m$, $L \propto \left(\frac{\beta I_0}{S}\right)^{K-N_r}$. If we let $\beta=\frac{S}{N_0}$, $L \propto \left(\frac{I_0}{N_0}\right)^{K-N_r}$. The interference-to-noise power ratio is a key parameter that dictates the user requirement. For $S=I_0$, we have: $L \propto SNR^{K-N_r}$.\\
Thus, each transmitter provides an outage capacity of $\log(1+\beta)$ with TOP $P_{\texttt{out,UB}}$. The sum of outage capacities of all $K$ transmitter is given by
\begin{eqnarray}
C_{\texttt{sum}, \texttt{Complex}} =  K\log(1+\beta)\label{Eq:Convsumcapacity}
\end{eqnarray}
This holds as long as $L$ satisfies (\ref{Eq:Lexact}). Thus, the proposed framework enables the transmitter to schedule a user who is nearly free of interference when the number of active users $L$ satisfies the stated requirement. It is important to note that the sum outage capacity grows linearly with the number of transmitters even in the case of single receiver antenna. Multiple receiver antennas play an important role here mainly in reducing the user requirement.
\section{Multi-user Spatial Multiplexing}\label{sec:3}
Next, we consider a generalized system model with $K$ transmitters each with $N_t$ transmit antennas. In every scheduling epoch, every transmitter serves $N_t$ users simultaneously with \emph{single data stream} allocated per user. Throughout this paper we assume that $L \ge N_t$. The baseband received signal for the user with index $l$ that is served by a given transmitter is represented as
\begin{eqnarray}\label{Eq:sysModel-1}
\by_{l}(k) =\sqrt{\frac{S}{N_t}}  \bH_{l} \bx(k) + \sum_{i=1}^{K-1} \sqrt{\frac{I_0}{N_t}} \bG_{i,l} \bx_{i}(k) +\bn_{l}(k), \quad \quad l=1,2,..L.
\end{eqnarray}
The total signal power is equally divided among the $N_t$ data streams. Here, $\bH_{l}$ and $\bG_{i,l}$ are modeled as multivariate circularly symmetric complex Gaussian random matrices having i.i.d. elements with $\textsf{E}[\bH_{l}]=\textsf{E}[ \bG_{i,l}]=0$ and $\textsf{E}[\bH_{l}
\bH^{\dagger}_{l}]=\textsf{E}[ \bG_{i,l}\bG^{\dagger}_{i,l}]=\bI$. The modulation vectors $\bx(k) \cong [x_{1}(k), x_{2}(k),..,x_{N_t}(k)]^{'}$ and $\bx_{i}(k) \cong [x_{i,1}(k), x_{i,2}(k),..,x_{i,N_t}(k)]^{'}$ are assumed to be complex-valued vectors whose elements are i.i.d. complex Gaussian random variables with zero mean, and unit variance, respectively.\\
In the proposed framework each user is restricted to receive a single complex-valued data stream from its transmitter. Therefore, every user can potentially receive data from any one of $N_t$ antennas of its own transmitter. We further propose that each user reports the post processing SINR of an MMSE receiver corresponding to all the $N_t$ data streams back to the transmitter. Let $i$ denote the index of the antenna through which the data is transmitted for a particular user. This index is referred to as the stream index (SI). Considering the symmetric channel with $S=I_0=1$, the signal model for detecting $i$th data stream can be represented as
\begin{eqnarray}\label{Eq:sysModel-1a}
\by_{l}(k) =\sqrt{\frac{1}{N_t}} \left[\underbrace{ \bh_{i,l} x_{i}(k)}_{\texttt{Desired Signal}} + \underbrace{\sum_{j \neq i}  \bh_{j,l} x_{j}(k)}_{\texttt{Self interference}} + \underbrace{\sum_{i=1}^{K-1} \bG_{i,l} \bx_{i}(k)}_{\texttt{Other cell interference}} \right] +\bn_{l}(k), \quad i=1,2,..,N_t
\end{eqnarray}
where $\bh_{p,l}$ is the channel vector of $p$th data stream with length $N_r \times 1$.  In detecting $i$th data stream, the remaining $N_t-1$ data streams transmitted by its own transmitter appear as \emph{self-interference}, in addition to \emph{other cell interference } contributed by the  $(K-1)N_t$ data streams transmitted by the $K-1$ co-channel transmitters. Thus a total of $KN_t-1$ data streams cause interference to the desired signal. The user determines the post-processing SINR of an MMSE receiver for the $i$th data stream as
\begin{eqnarray}
\gamma_{i,l}=  \bh^{\dagger}_{i,l} \bR^{-1}_{i,l} \bh_{i,l}, \quad i=1,2,..,N_t
\end{eqnarray}
where
$
\bR_{i,l}= \sum_{j \neq i}  \bh_{j,l} \bh^{\dagger}_{j,l}+\sum_{i=1}^{K-1}  \bG_{i,l} \bG^{\dagger}_{i,l} + N_t N_0 \bI
$
is the total noise-plus-interference covariance matrix. We are particularly interested in the case when the ICM has full rank. This happens when $KN_t-1 \ge N_r$ i.e., the number of interfering data streams is greater than or equal to the receiver antenna array size. The user scaling rules for the MU SM can be obtained by extending the user scaling results obtained for the case of SST with complex-valued encoding. Using a sub-optimum sequential max-SINR scheduler that is described in Appendix A, the sum of outage capacities of all $N_t$ streams for all  $K$ transmitters is shown to be
\begin{eqnarray}
C_{\texttt{sum, MU SM}}= KN_t \log(1+\beta)
\end{eqnarray}
with each stream meeting the outage probability constraint given in (\ref{Eq:muexact}). For large values of  $\beta+m$, and when $L >> N_t$,
$
L \approx e^{\beta N_t N_{0}} \ln (P^{-1}_{\texttt{out},\texttt{UB}}) \frac{\left(\beta+ m\right)^{KN_t-N_r}}{a(\bar{k}_0) (\bar{k}_0)!}.
$
For, $\beta \cong \frac{1}{N_0}$, using suitable approximations we can show that the number of active users required to meet certain target per stream outage probability is proportional to $SNR^{KN_t-N_r}$. To achieve interference free performance, SM requires a significantly higher number of active users. In the following we propose a real-valued transmission scheme that reduces the user requirement. First, we analyze the performance of this method for the case of SST followed by a generalization to the case of MU SM employing real-valued encoding.

\section{SST with Real-valued Encoding} \label{sec:4}
In the proposed system model, all the transmitters in the network transmit \emph{a single data
stream} using \emph{real-valued} modulation alphabets. The receiver at each user collects the real and imaginary parts of the multi-antenna receiver to obtain a virtual antenna
array of size $2N_r$. The receiver filters the real and imaginary parts of the received signal using a WL MMSE filter for data detection. The post-processing SINR of the receiver is reported back to the transmitter using a feedback channel. Scheduling and MCS allocation is done based on the post-SINR of the WL MMSE. We evaluate the
TOP for this type of encoding. First, we begin with the system model for transmission of real-valued modulation symbols
\begin{eqnarray}
\by_{l}(k) =\sqrt{S}  \bh_{l} \bar{x}_{l}(k) + \sum_{i=1}^{K-1} \sqrt{I_0} \bg_{i,l} \bar{x}_{i,l}(k) +\bn_{l}(k),
\end{eqnarray}
where $\bar{x}_{l}(k)$ and $\bar{x}_{i,l}(k)$ are real-valued modulation alphabets of the desired signal and interference, respectively. The baseband receiver collects the real and
imaginary  parts of the complex-valued received signal for each antenna branch and collects the observations in a vector-format as: $ \tilde{\by}_{l}(k)
=\sqrt{S}  \tilde{\bh}_{l} \bar{x}_{l}(k) + \sum_{i=1}^{K-1} \sqrt{I_0} \tilde{\bg}_{i,l} \bar{x}_{i,l}(k) +\tilde{\bn}_{l}(k), \label{Eq:WLcase} $ where the notation $\tilde{\bx}=\left[\begin{array}{c}
                                  \texttt{real}(\bx) \\
                                  \texttt{imag}(\bx)
                                \end{array}\right]
$ denotes a vector with real and imaginary parts stacked in a column vector format. Here, $\tilde{\bh}_{l}$, $\tilde{\bg}_{i,l}$ contain the real and imaginary parts of the desired
and interfering channels, respectively, and $\tilde{\bn}_{l}(k)$ contains the real and imaginary parts of the noise samples. Since the individual elements of the complex-valued channel vectors $\bh_l$ are assumed to be i.i.d. circular complex Gaussian random variables, the real and imaginary parts of $\bh_l$ are also zero-mean, i.i.d. Gaussian. Therefore, $\tilde{\bh}_{l} \sim \textsf{N}(0,\frac{1}{2}\bI)$, where the notation
denotes a multivariate real Gaussian distribution with zero-mean and variance $\frac{1}{2}\bI$. Similarly, $\tilde{\bg}_{i,l} \sim
\textsf{N}(0,\frac{1}{2}\bI)$ for $i=1,2,..,K-1$. The real-valued modulation sequences $\bar{x}_{l}(k)$ and $\bar{x}_{i,l}(k)$, for each $i$, are assumed to be  i.i.d. real Gaussian random variables with zero-mean, unit variance, and statistically independent of each other.

\subsubsection*{Capacity scaling laws for real-encoding}
Using the results of Appendix B, the sum of outage capacities of $K$ transmitters employing real-valued encoding is given by
\begin{eqnarray}
C_{\texttt{sum}, \texttt{Real}} &=& \frac{K}{2}\log(1+\beta)  \label{Eq:PAMsumcapacity}
\end{eqnarray}
This result hold when: $L \propto \left( \frac{ I_0 \beta}{S}+m\right)^{\frac{K}{2}-N_r}$. For $\beta=\frac{S}{N_0}$, for $S=I_0$ and when $\frac{ I_0 \beta}{S} > m$, we have: $L \propto SNR^{\frac{K}{2}-N_r}$. Compared to SST with complex-valued encoding, the proposed real encoder requires significantly less number of users. The user reduction is achieved at the expense of a pre-log rate reduction by a factor of $\frac{1}{2}$. Numerical and simulation comparison shows that these two methods are optimal in different operating regimes.
\subsection{Generalized MU SM with real-encoding}
The results for SST with real-valued encoding can be generalized to MU employing real-valued encoding.  Following the analysis for the case of MU SM with complex-valued encoding, it can be shown that by spatial multiplexing $t$ real-valued data streams using $t$ antennas, we  get a sum outage capacity of
\begin{eqnarray}
C_{\texttt{sum, real MU SM}}= \frac{tK}{2} \log(1+\beta)
\end{eqnarray}
when the number of active users $L \propto \beta^{\frac{tK}{2}-N_r}$. The proof follows the same line of arguments used in the case of SM employing complex-valued encoding using a sequential max-SINR scheduler. While complex-valued MU SM offers a SM rate of $R=N_t$ where $N_t$ takes integer values, real-valued MU SM gives fractional multiplexing rates of $R=\frac{t}{2}$ which take values in steps of $0.5$. The real-valued encoder can be viewed as a generalized SM encoder. Using $t=2N_t$, we get the same user scaling results as that of complex-valued encoding. However, real encoding offer a wider range of multiplexing rates and therefore offers a finer trade-off between outage capacity and the number of active users. Simulation shows that use of real-valued encoding offers a performance that is either comparable to complex encoding or exceeds by a significant margin. Detailed results are given in section \ref{sec:5}.

\subsubsection*{SIA feasibility in non-Rayleigh fading channels}
We remark here that though we analyze the system performance for the important case of i.i.d. Rayleigh fading channels, the SIA gains can be obtained in channels with arbitrary type of fading as long as the interference channel vectors exhibit linear dependency. In channels with full magnitude correlation between receiver antenna branches, the SIA phenomenon occurs as long as the phase vectors of the channel takes random values. In case of Rician channels, the channel has a line-of-sight (LOS) term and a Rayleigh fading component. The SIA feasibility in case of LOS channels needs careful attention.
When the signal and interference channel have strictly LOS component, then: a) individual interferers often arrive at different angles b) inter-antenna spacing causes a phase difference among the channel states of different antennas, and these phase differences take distinct values for different interferers arriving at different angles. Essentially, any two interference channel vectors become linearly independent as long as their angles of arrival are sufficiently distinct. Therefore, the probability of occurrence of SIA increases if the signal and interferers always arrive at distinct angles. One needs to carefully study the performance of opportunistic scheduling for Rician case using more complex channel models.\\
In case of real encoding, the channel vector contains the real and imaginary parts of the complex-valued channel. Consider the special case of single receiver antenna. The channel gain between a given transmitter receiver pair almost always takes complex-values independent of whether the channel has LOS, Rayleigh, or Rician distribution. Even in the case of LOS channel, the phase angles of channels of signal and interferers are statistically independent. Consequently, the interference channel vectors (that contain real and imaginary parts of a complex scalar) take values such that SIA occurs with high probability when the number of users is sufficiently large. However, for the case of real-encoding with multiple antennas, the full benefit can be realized when channel phase states are statistically independent among antenna branches, and among interferers.

\section{Results and Discussion}\label{sec:5}
\subsection{Comparison of SST with real and complex encoding }
\begin{figure}
    \centerline{
        \epsfig{figure=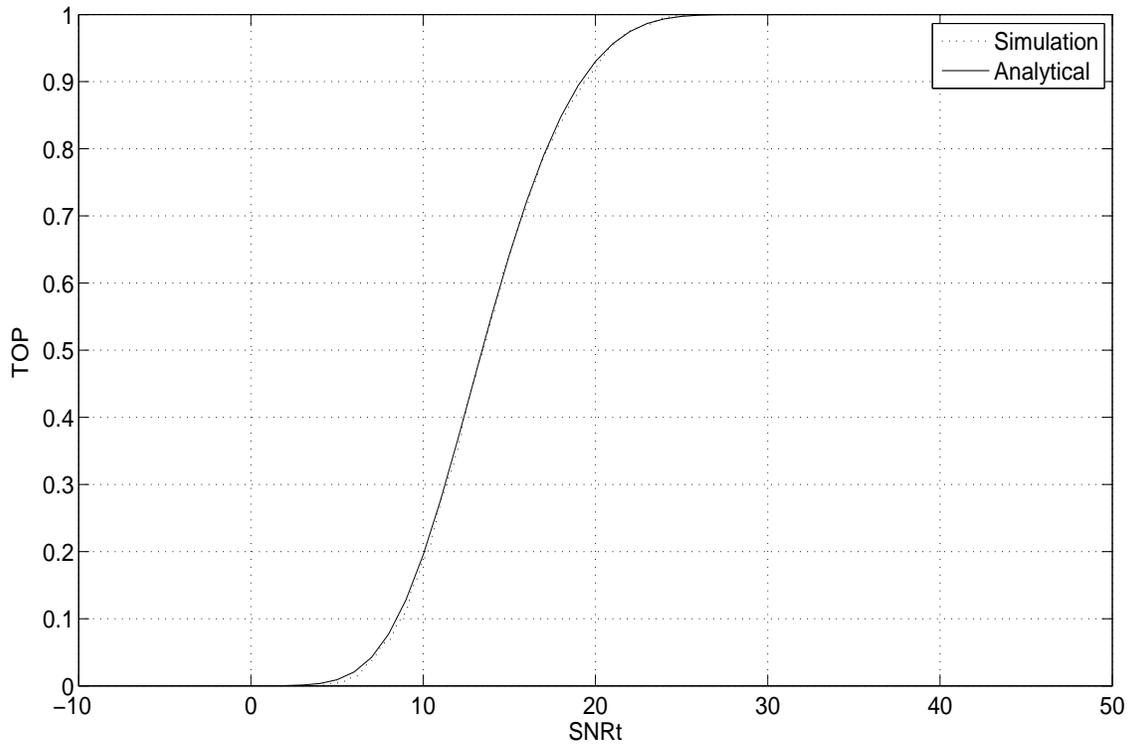, height=10cm, width=15.0cm}
    }
        \vspace*{0.0cm}  \caption{TOP for complex-valued encoding, $N_r$=2, $K$=3, SNR=20 dB, $L$=10} \label{fig:0}

\end{figure}

\begin{figure}
    \centerline{
        \epsfig{figure=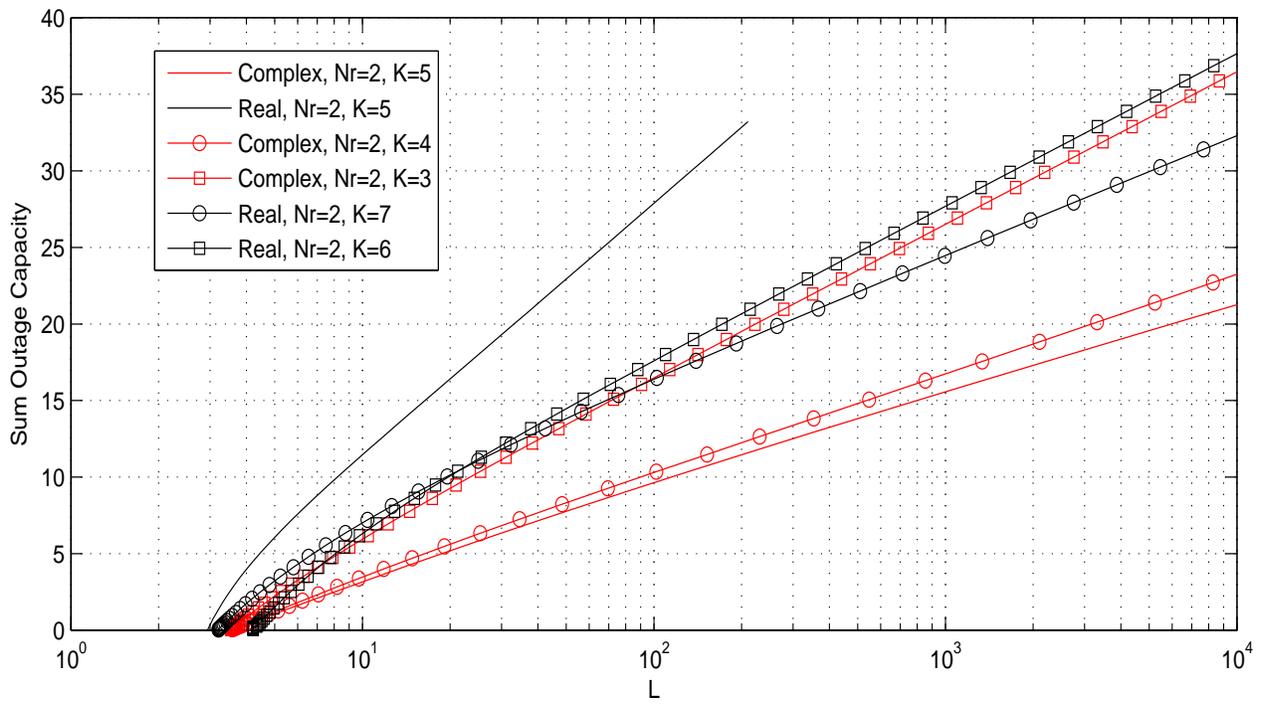,  height=10cm, width=20.0cm }
    }
        \vspace*{0.0cm}  \caption{$Nr$=2, Pout=0.2, SNR=20 dB, Sum outage capacity Vs L} \label{fig:1c}

\end{figure}

\begin{figure}
    \centerline{
        \epsfig{figure=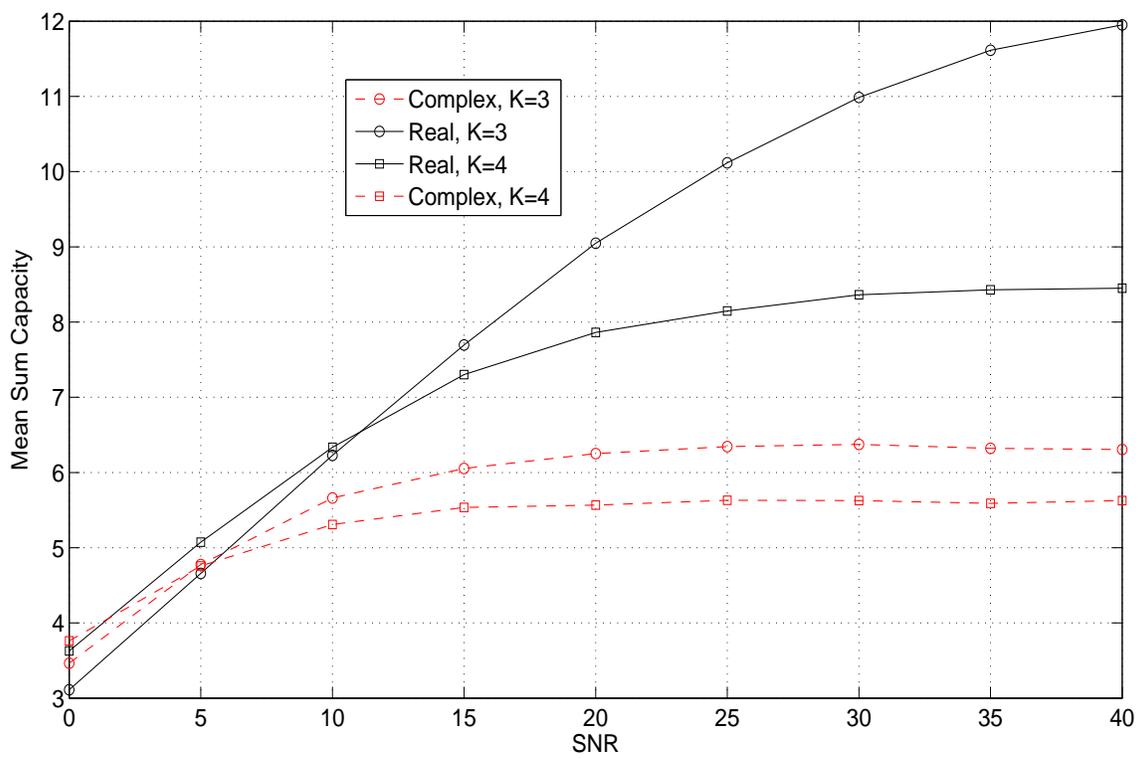, height=10cm, width=15.0cm}
    }
        \vspace*{0.0cm}  \caption{Mean Capacity for  $N_r=1$, $L=10$} \label{fig:2}

\end{figure}

\begin{figure}
    \centerline{
        \epsfig{figure=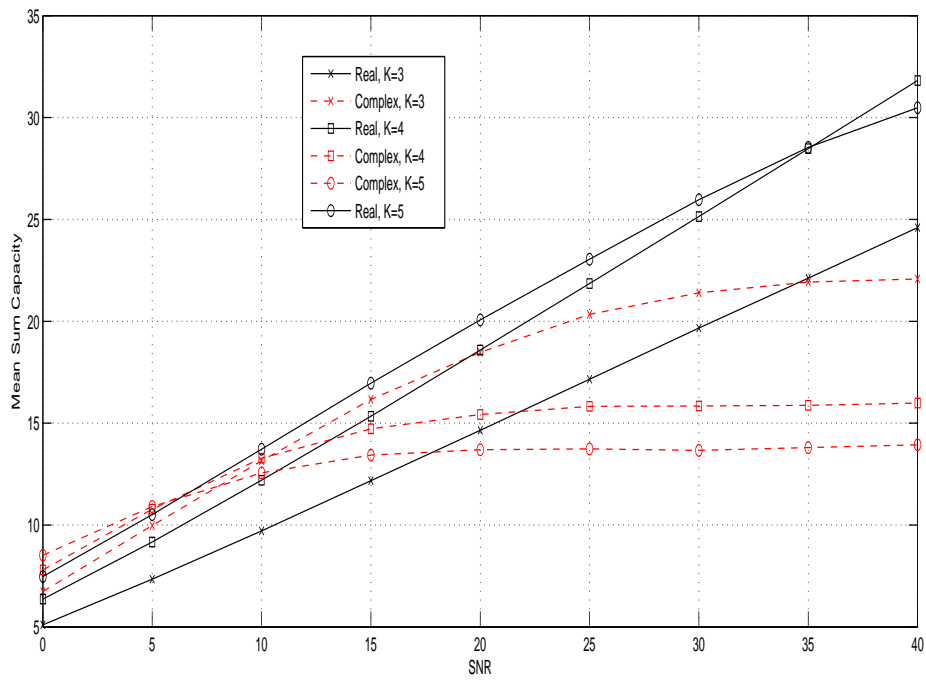, height=10cm, width=15.0cm}
    }
        \vspace*{0.0cm}  \caption{Mean Capacity for  $N_r=2$, $L=50$} \label{fig:5}

\end{figure}

Throughout the rest of the section, we assume that $S=I_0=1$. The target SNR ($\beta$) is denoted as: $SNR_t$. In Fig \ref{fig:0}, the analytically obtained TOP results are compared with simulation results for the case of complex encoding. The legend exact UB refers the exact TOP UB and approx refers to the various approximations used in arriving at a closed-form expression for the TOP. Fig \ref{fig:0} shows that the UB on TOP is extremely tight for complex-valued encoding.  Fig \ref{fig:1} shows that the UB on TOP given by (\ref{Eq:outagewlf}) in Appendix B 2 for real-valued encoding deviates slightly for low values of SNR however it becomes a tight bound for moderate to high SNR values. Additionally, the Q-function based approximation given in  Appendix B 4 is fairly accurate for even values of $K$ for real-valued signaling case. In Fig \ref{fig:1a}, it is shown that the TOP approximations given in Appendix B 5.1 for real-valued encoding are tight for $K=2N_r+1$. \\
In Fig \ref{fig:1c}, we plot sum outage capacity as a function of number of active users for the case of $N_r=2$. We consider the important case where the number of interferers either equals or exceeds the receiver array size.  Results show that real-valued encoding with $K=5$ provides a significantly higher sum outage capacity compared to other feasible configurations involving real/complex encoding.\\
Next, we discuss the mean sum capacities of proposed techniques. The results are obtained using simulation where results are averaged over 1000 channel realizations.\\
Fig \ref{fig:2}, show the results with single receiver antenna for $L$=10. The performance is quite remarkable since we are able to get fairly high capacities using a single receiver antenna. Complex-valued signaling outperforms real-valued signaling in low to medium SNR range. At high SNR, real-valued signaling performs significantly better as mean sum capacities of complex-valued signaling reach saturation.\\
Fig \ref{fig:5} shows results for two receiver antenna case for $L=50$. For real-valued signaling, mean sum capacity grows linearly with SNR when $K<5$. For $K=3$, complex-valued signaling shows near linear growth and this mode outperforms real-valued signaling. However, for $K>3$, the gain of real-valued signaling over complex-modulation is substantially high. In Tables II and III, the mean sum capacity results for real and complex-valued encoding methods are tabulated for the case of $N_r=1$ and $N_r=2$, respectively. In each column, the method with highest mean sum capacity is highlighted.
\subsection{Performance comparison between SST, SU and MU SM}\label{singlestVsmulti}
\begin{figure}
    \centerline{
        \epsfig{figure=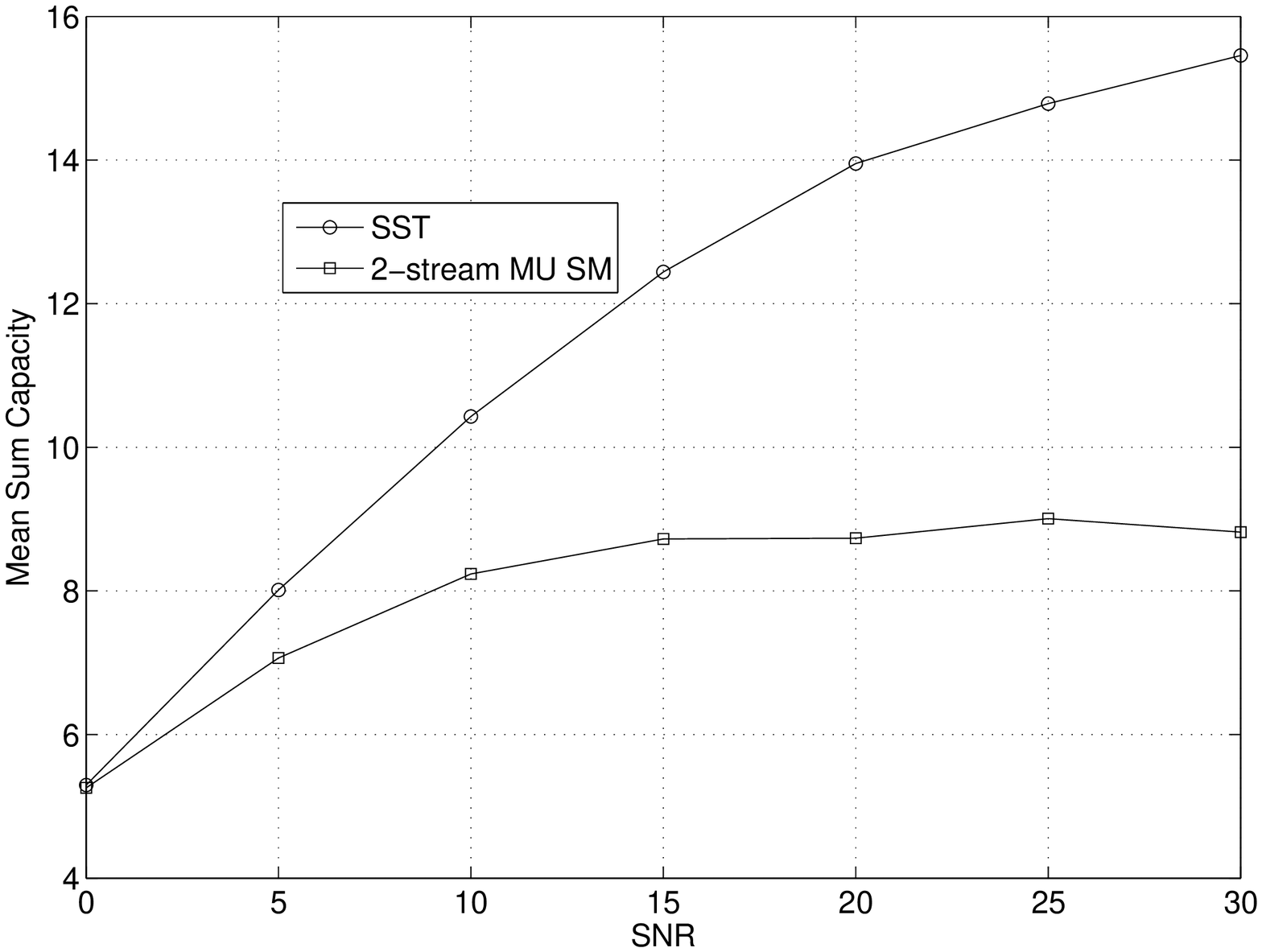, height=10cm, width=15.0cm  }
    }
        \vspace*{0.0cm}  \caption{Mean Sum Capacity Comparison of SST and SM for $N_r=2$, $L=10$, $K=3$} \label{fig:6}
\end{figure}

\begin{figure}
    \centerline{
        \epsfig{figure=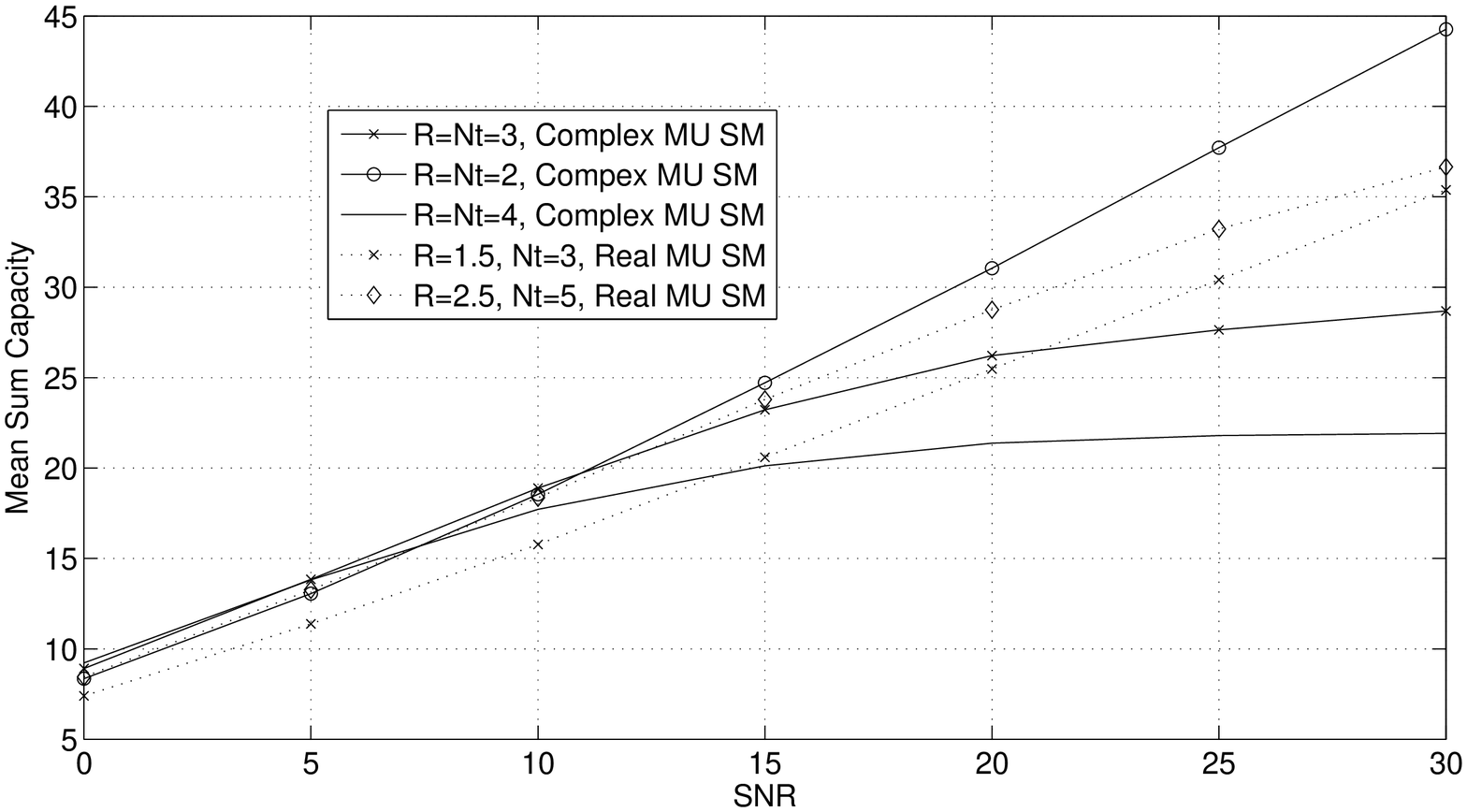, height=10cm, width=15.0cm  }
    }
        \vspace*{0.0cm}  \caption{Mean Sum Capacity Comparison of SST and SM for $N_r=4$, $L=50$, $K=2$} \label{fig:9}
\end{figure}

\begin{figure}
    \centerline{
        \epsfig{figure=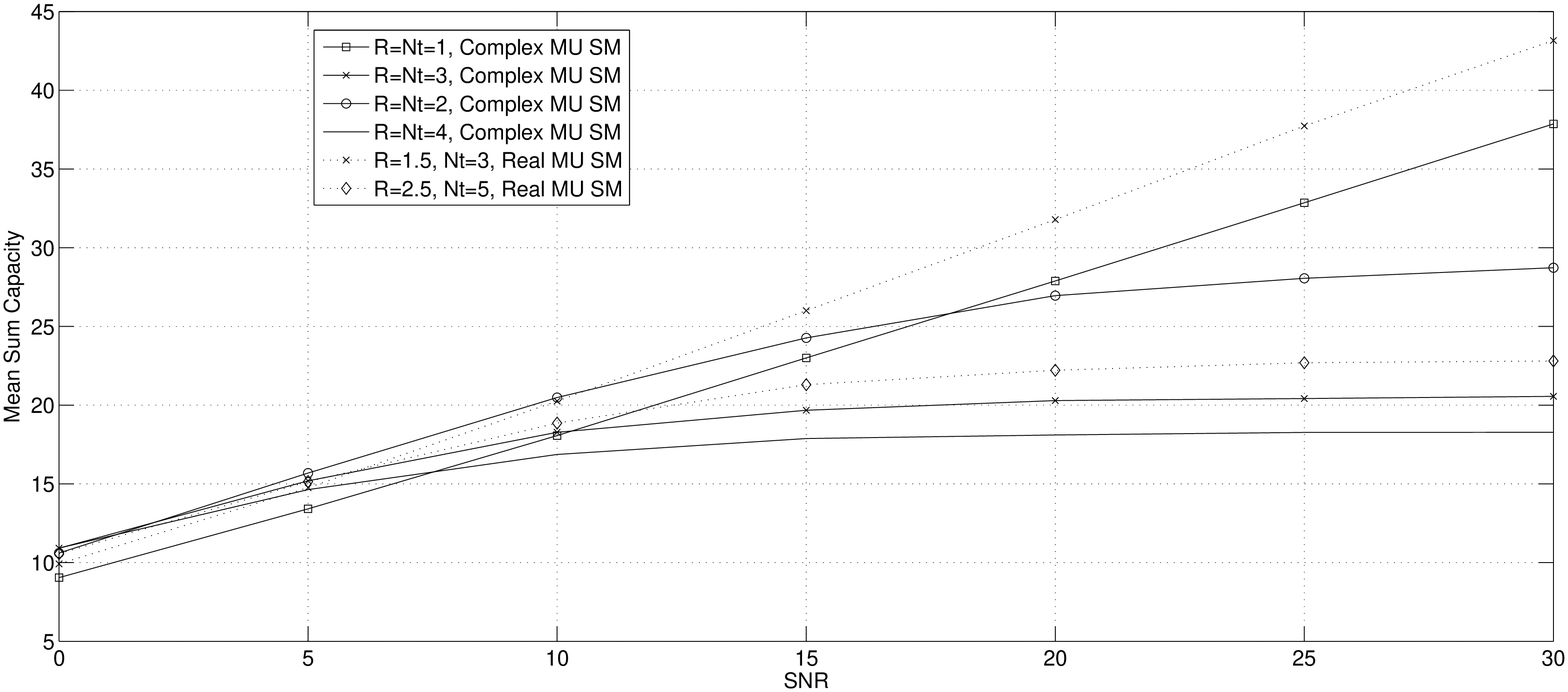, height=10cm,, width=15.0cm  }
    }
        \vspace*{0.0cm}  \caption{Mean Sum Capacity Comparison of SST and SM for  $N_r=4$, $L=50$, $K=3$} \label{fig:10}
\end{figure}

\begin{figure}
    \centerline{
        \epsfig{figure=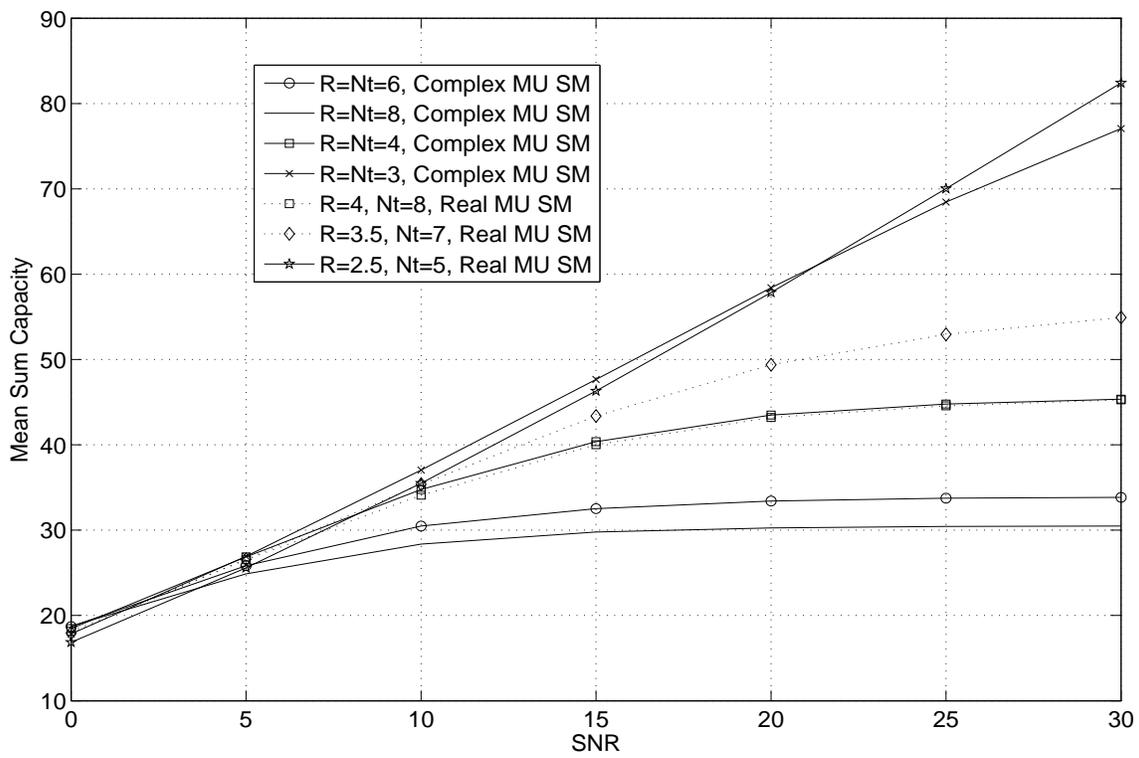, height=10cm, width=15.0cm  }
    }
        \vspace*{0.0cm}  \caption{Mean Sum Capacity Comparison $N_r=8$, $L=100$, $K=3$} \label{fig:2f}

\end{figure}

In all the figures the SM rate is defined as: $R=\frac{N_t}{2}$ for real encoding and $R=N_t$ for complex encoding. Fig \ref{fig:6}, shows results for $N_r=2$, $L=10$ and $K=3$. We see that SST with complex-valued encoding outperforms 2-stream MU SM with complex encoding. \\
Fig \ref{fig:9} shows performance results for the case with $N_r=4$, 50 active users, and $K=2$. Limiting to 2-streams using complex encoding gives better performance compared to the rest of the cases. However with $K=3$, results of Fig \ref{fig:10} shows that 3-stream MU SM with real encoding with a rate of $R=1.5$ gives significant gain over the case of $R=2$ which uses complex-valued encoding.\\
In Fig \ref{fig:2f}, we compare sum capacity results for $N_r=8$, with $K=3$, $L=100$. For this case, a rate of 2.5 or 3 outperforms all other modes. In particular, 5-stream MU SM with real encoding with $R=2.5$ outperforms 3-stream MU SM which employs complex-valued signaling at high SNR. However, the performance for both cases is comparable in the medium SNR range. Also note that for a SM rate of $R=4$, MU SM with real and complex encoding have comparable sum capacity.\\
\textbf{\texttt{Remarks}}
\begin{itemize}
  \item We show that MU SM with real encoding offers a higher sum capacity compared complex encoding in certain cases. This is accomplished by increasing the number streams/antennas at each transmitter. The number of used antennas can be reduced further using a combination of real and complex encoding. For example, let us assume that each transmitter transmits $m$ complex-valued data streams using $m$ antennas, and  $N_t-m$ real-valued data streams using the remaining $N_t-m$ antennas. Thus, the BS serves $t=2m+N_t-m=N_t+m$ users using $N_t$ antennas. The total spatial multiplexing rate for each transmitter is: $R=\frac{N_t+m}{2}$, $m$ takes values in the range $[0,N_t]$ where the extreme values represent real only, or complex-only encoders. For $m\in [0,N_t]$, we get multiplexing rates in the range $ [\frac{N_t}{2},\frac{N_t+1}{2},..,\frac{2N_t-1}{2}, N_t]$ using a suitable mix of real and complex modulations. For example, to get a SM rate of 2.5, real encoding uses 5 real-valued streams using $N_t=5$. For this mixed encoding case, we use a total of 3 antennas where the first two antennas employ complex-valued encoding and the third antenna employs real encoding. The total number of streams is 5 using 3 transmit antennas. For this case, the receiver for each user uses WL-MMSE processing as in the case of real encoding. Simulation shows that this type of encoding provides a performance similar to the case of real only encoding. Detailed results are not shown due to space limitations.
      \item In cellular networks IA is applicable where interference is high, that is for a cell edge user. For a user at the edge of the cell, the distances from the active BS and the interfering BSs are comparable. As the BSs are assumed to use equal transmit power, the interference power levels are approximately equal. In this situation, SIC model is justified. This model is used in \cite{Inkyu:12},\cite{Jung:12},\cite{Paul:IA} as well.
          \item  Recently, \cite{Kuchi:ICC2013} presented a tractable approach to coverage and rate evaluation for MIMO cellular networks using stochastic geometry methods \cite{Andrews:Sept2010}. Using ZF receivers, it is shown that that SM degrades the rate for a notable percentage of users compared to single stream transmission. For the case of two receiver antennas, the increase in mean rate of SM is shown to be modest compared to SST, while SST is shown to provide a gain in rate for cell edge users. However, for higher antenna configurations, reducing the SM rate to a value less than the maximum allowed rate is shown to offer an overall increase in the system performance. In this paper, we observe a similar trend using a SIC model when opportunistic scheduling is combined with MMSE interference suppression. Therefore, the performance of the proposed encoding methods needs to be investigated further in both conventional and Heterogeneous cellular networks employing opportunistic scheduling \cite{Kuchi:GC12}.
\end{itemize}
\section{Conclusions}\label{sec:7}
This paper highlights the spatial interference alignment phenomenon that naturally occurs in multi-user systems employing opportunistic scheduling. For the case of symmetric interference channel, closed-form expressions for outage capacity and capacity scaling laws with number of users is obtained for a single stream and multi-stream SM systems employing real or complex-encoding. \\

We show that SST methods employing real and complex-valued encoding methods have distinct sum capacities and the two methods are optimal in different operating regimes. In an $N_r$ receiver antenna system employing SST with complex-encoding, use of opportunistic scheduling based on post-SINR of MMSE receiver enables mitigation of more than $N_r-1$ interferers. For $K=N_r+1$, the required number of users scale linearly with SNR. For the case of real-encoding, mitigation of more than $2N_r-1$ interferers is feasible. Though SST with real-encoding reduces the peak rate by a factor of two, the overall sum capacity exceeds that of complex-valued encoding in certain cases. For the special case of $K=2N_r+1$, the required number of users for real encoding scales with $\sqrt{SNR}$ where it provides a higher sum outage capacity compared to complex encoding.\\

We generalized the SST with complex/real encoding to MU SM case. The real/complex encoding with a spatial multiplexing rate of $R$, we  get a sum outage capacity of $KR \log(1+SNR)$ when $L \propto SNR^{KR-N_r}$. With $N_t$ antennas, $R=\frac{N_t}{2}$ for real encoding and $R=N_t$ for complex encoding. We show that the generalized MU SM encoder with real-valued modulation provides a performance that is either comparable, or significantly higher than that of complex encoding. The additional gain owes to the fact that real-valued MU SM offers a wider range of multiplexing rates and offers a finer trade-off between achievable capacity and user requirement.\\

In systems with significant amount of interference, a reduction in SM rate is shown to have a beneficial effect of increasing the overall sum capacity. With two receiver antennas at the user, SST mode employing complex encoding and opportunistic scheduling outperforms SM. With four receiver antennas, reducing the SM rate to either 1.5 or 2 is preferable over full rate transmission i.e, $R=4$. The proposed encoding methods can be used to improve the cell edge user rate in cellular systems.

\bibliographystyle{IEEEtran} 
\bibliography{IEEEabrv,IEEEexample}

\begin{thebibliography}{10}
\providecommand{\url}[1]{#1}
\def\UrlFont{\rmfamily}
\providecommand{\newblock}{\relax}
\providecommand{\bibinfo}[2]{#2}
\providecommand\BIBentrySTDinterwordspacing{\spaceskip=0pt\relax}
\providecommand\BIBentryALTinterwordstretchfactor{4}
\providecommand\BIBentryALTinterwordspacing{\spaceskip=\fontdimen2\font plus
\BIBentryALTinterwordstretchfactor\fontdimen3\font minus
  \fontdimen4\font\relax}
\providecommand\BIBforeignlanguage[2]{{%
\expandafter\ifx\csname l@#1\endcsname\relax
\typeout{** WARNING: IEEEtran.bst: No hyphenation pattern has been}%
\typeout{** loaded for the language `#1'. Using the pattern for}%
\typeout{** the default language instead.}%
\else
\language=\csname l@#1\endcsname
\fi
#2}}

\bibitem{Cadambe:2008}
V.~Cadambe and S.~Jafar, ``Interference alignment and degrees of freedom of the
  {K}-user interference channel,'' \emph{{IEEE} Trans. Inform. Theory},
  vol.~54, pp. 3425--41, Aug. 2008.

\bibitem{Maddah:2008}
M.~A. Maddah-Ali and A.~S. Motahari, ``Communication over {MIMO} {X} channels:
  {I}nterference alignment, decomposition, and performance analysis,''
  \emph{{IEEE} Trans. Inform. Theory}, vol.~15, pp. 3457--70, Aug. 2008.

\bibitem{Jafar:2009}
B.Nazer, M.~S. Jafar, and S.~Vishwanath, ``Ergodic interference alignment,''
  \emph{{IEEE} Trans. Inform. Theory}, Submitted for publication.

\bibitem{Slock:IA}
F.~Negro, S.P.Shenoy, D.~Slock, and I.~Ghauri, ``Interference alignment limits
  for {K}-user frequency-flat {MIMO} interference channels,'' in \emph{17th
  European Signal Porcessing Conference}, Aug. 2009.

\bibitem{Motahari:2009}
A.~Ghasemi, A.~S. Motahari, and A.~K. Khandani, ``Interference alignment for
  the {K} user {MIMO} interference channel,'' \emph{{IEEE} Trans. Inform.
  Theory}, vol.~15, 2009.

\bibitem{Cadambe:2010}
V.~R. Cadambe and S.~A. Jafar, ``Interference alignment with asymmetric complex
  signaling: Settling the høst madsen nosratinia conjecture,'' \emph{{IEEE}
  Trans. Inform. Theory}, vol.~15, pp. 462--465, 2010.

\bibitem{Inkyu:12}
H.Y.Shin, S.~Park, H.~Park, and I.~Lee, ``A new approach of interference
  alignment through asymmetric complex signaling and multiuser diversity,''
  \emph{{IEEE} Trans. Wireless Commun.}, Mar. 2012.

\bibitem{Knopp:ICC95}
R.~Knopp and P.~Humblet, ``Information capacity and power control in
  single-cell multiuser communications,'' in \emph{ICC}, 1995, pp. 331--335.

\bibitem{Tse:2002}
P.~Viswanath, D.~Tse, and R.~Laroia, ``Opportunistic beamforming using dumb
  antennas,'' \emph{{IEEE} Trans. Inform. Theory}, vol.~47, pp. 1277--94, Mar.
  2002.

\bibitem{Ho}
Z.~Ho and E.~Jorswieck, ``Improper {G}aussian signaling on the two-user {SISO}
  interference channel,'' \emph{{IEEE} Trans. Wireless Commun.}, pp.
  3194--3203, 2012.

\bibitem{Jung:12}
B.C.Jung, D.~Park, and W.Y.Shin, ``Opportunistic interference mitigation
  achieves optimal degrees-of-freedom in wireless multi-cell uplink networks,''
  \emph{{IEEE} Trans. Commun.}, Feb. 2012.

\bibitem{Paul:IA}
\BIBentryALTinterwordspacing
H.~Yang, W.Y.Shin, B.C.Jung, and A.~Paulraj, ``Opportunistic interference
  alignment for {MIMO} interfering multiple-access channels.'' [Online].
  Available: \url{http://arxiv.org/abs/1302.5280}
\BIBentrySTDinterwordspacing

\bibitem{Telatar:June95}
E.Telatar, ``Capacity of multi-antenna gaussian channels,'' in \emph{ATT-Bell
  Labs Internal Tech. Memo}, 1995, pp. 585--595.

\bibitem{Foschini:Mar98}
G.J.Foschini and M.J.Gans, ``On limits of wireless communication in a fading
  environment when using multiple antennas,'' in \emph{Wireless personal
  communications}, Mar. 1998, pp. 311--335.

\bibitem{Greedy:Oct05}
Q.~Spencer, A.~Swindlehurst, and M.~Haardt, ``Zero-forcing methods for downlink
  spatial multiplexing in multi-user {MIMO} channels,'' \emph{{IEEE} Trans.
  Commun.}, vol.~53, pp. 3857--3868, Oct. 2005.

\bibitem{Peel:Jan05}
C.~Peel, B.~Hochwald, and A.~Swindlehurst, ``A vector-perturbation technique
  for near-capacity multiantenna multiuser communication-part {I}: {C}hannel
  inversion and regularization,'' vol.~53, pp. 195--202, Jan. 2005.

\bibitem{Peel:Mar05}
------, ``A vector-perturbation technique for near-capacity multiantenna
  multiuser communication-part {I}: {C}hannel inversion and regularization,''
  vol.~53, pp. 537--544, Mar. 2005.

\bibitem{picinbono:95}
B.~Picinbono and P.~Chevalier, ``Widely linear estimation with complex data,''
  \emph{{IEEE} Trans. Signal Processing}, vol.~43, pp. 2030--2033, Aug. 1995.

\bibitem{Gerstacker:Sept03}
W.~H. Gerstacker, R.~Schober, and A.~Lampe, ``Receivers with widely linear
  processing for frequency-selective channels,'' \emph{{IEEE} Trans. Commun.},
  vol.~51, pp. 1512--22, 2003.

\bibitem{Chevaliar:Mar2006}
P.~Chevalier and F.Pipon, ``New insights into optimal widely linear array
  receivers for demodulation of {BPSK}, {MSK}, and {GMSK} corrupted by
  noncircular interferers-{A}pplication to {SAIC},'' \emph{{IEEE} Trans. Signal
  Processing}, vol.~54, pp. 870--883, Mar. 2006.

\bibitem{Kuchi:Jan09}
K.Kuchi and V.K.Prabhu, ``Performance evaluation for widely linear demodulation
  of {PAM/QAM} signals in the presence of rayleigh fading and co-channel
  interference,'' \emph{{IEEE} Trans. Commun.}, vol.~57, Jan. 2009.

\bibitem{Cioffi:Stanford}
\BIBentryALTinterwordspacing
{J. Cioffi}, \emph{EE:379 Stanford Class Notes}. [Online]. Available:
  \url{http://www.stanford.edu/class/ee379a/}
\BIBentrySTDinterwordspacing

\bibitem{Edelman}
\BIBentryALTinterwordspacing
A.~Edelman, ``Eigenvalues and condition numbers of random matrices,'' Ph.D.
  dissertation, Massachusetts Institute of Technology, May 1989. [Online].
  Available: \url{http://www-math.mit.edu/~edelman/Edelman/publications.htm}
\BIBentrySTDinterwordspacing

\bibitem{Chiani:Feb03}
M.~Chiani, M.Z.Win, A.~Zanella, R.K.Malik, and J.H.Winters, ``Bounds and
  approximation for optimum combining of signals in the presence of multiple
  cochannel interferers,'' \emph{{IEEE} Trans. Commun.}, vol.~51, pp. 296--307,
  Feb. 2003.

\bibitem{Wishart:Natraj}
T.~Ratnarajah, R.~Vaillancourt, and M.~Alvo, ``Eigenvalues and condition
  numbers of complex random matrices,'' \emph{{SIAM} Journal on Matrix Analysis
  and Applications}, 2004.

\bibitem{Kwan:07}
R.~Kwan, C.~Leung, and P.~Ho, ``Distribution of ordered eigenvalues of wishart
  matrices,'' \emph{{IEEE} Commun. Lett.}, vol.~43, Mar. 2007.

\bibitem{Chinai:April09}
A.~Zanella and M.~W. M.Chiani, ``On the marginal distribution of the
  eigenvalues of wishart matrices,'' \emph{{IEEE} Trans. Commun.}, vol.~57, pp.
  1050--60, Apr. 2009.

\bibitem{Chinai:12}
A.~Zanella and M.~Chiani, ``Reduced complexity power allocation strategies for
  {MIMO} systems with singular value decomposition,'' \emph{{IEEE} Trans. Veh.
  Technol.}, vol.~61, Nov. 2012.

\bibitem{book:Murihead}
R.~Murihead, \emph{Aspects of Multivariate Statistical Theory}.\hskip 1em plus
  0.5em minus 0.4em\relax Wiley Series in Probability and Mathematical
  Statistics, 1982.

\bibitem{Khatri:64}
C.G.Khari, ``Distribution of the largest or the smallest characteristic root
  under null hyphotesis concerning complex multivariate normal populations,''
  \emph{Ann. Math. Stat.}, vol.~35, 1964.

\bibitem{Kuchi:ICC2013}
T.~Sreejith, K.Kuchi, A.~Krishnaswami, and R.~Ganti, ``Coverage and rate in
  cellular networks with multi-user spatial multiplexing,'' in \emph{ICC},
  2013.

\bibitem{Andrews:Sept2010}
J.~G. Andrews, F.~Baccelli, and R.~K. Ganti, ``A tractable approach to coverage
  and rate in cellular networks,'' \emph{{IEEE} Trans. Commun.}, 2010.

\bibitem{Kuchi:GC12}
R.~Ganti and K.Kuchi, ``{SINR} order statisitics in {OFDMA},'' in
  \emph{Globecom}, 2012.

\bibitem{Chiani:July03}
M.~Chiani, D.Dardari, and M.K.Simon, ``New exponential bounds and
  approximations for the computation of error probability in fading channels,''
  \emph{{IEEE} Trans. Wireless Commun.}, vol.~2, pp. 840--45, 2003.

\end{thebibliography}

\section*{Appendix A \\ Capacity scaling laws for MU MIMO using sequential max-SINR scheduler}\label{Appendix2}
\subsection*{Appendix A 1 \\ Sequential Max-SINR Scheduler}\label{Appendix2:seqsch}
In a given scheduling epoch, the transmitter determines a group of $N_t$ users from the available set of $L$ users that provides maximum sum capacity among all feasible groups. The computational complexity of the search algorithm which determines the optimum group is quite large for large values of $(L,N_t)$. We propose a sub-optimum algorithm with low implementation complexity and good performance. In the proposed method, the transmitter determines the user to be scheduled for each stream index using a \emph{sequential} max-SINR scheduler. More specifically, let $U_1 \cong \{ \gamma_{1,1}, \gamma_{1,2},..,\gamma_{1,L} \}$ denote the channel quality information (CQI) metrics reported by all $L$ available users for the first stream index. For $i=1$, the scheduler first selects a user using the following rule:
$
\gamma_{1,l^{*}(1)}= \texttt{max} \quad U_1
$
where $l^{*}(1)$ is the index of the user whose CQI is maximum. Let $U_2 \cong \{ \gamma_{2,1}, \gamma_{2,2},..,\gamma_{2,L} \}$ denote the CQI metrics reported by all $L$ available users for stream index 2. We determine the scheduling decision for SI $i=2$ using the new set $\bar{U}_2$ which is obtained by excluding the CQI of the previously scheduled user from the set $U_2$ i.e., $\bar{U}_2=U_2-\{ \gamma_{2,l^{*}(1)} \}$. The scheduler selects a user as
$
\gamma_{2,l^{*}(2)}= \texttt{max} \quad \bar{U}_2.
$
Generalizing in this manner, we have: $U_{i} \cong \{ \gamma_{i,1}, \gamma_{i,2},..,\gamma_{i,L} \}$ denote the CQI metrics reported by all $L$ available users for $i$th stream index. Let, $\bar{U}_i \cong U_{i}-\{\gamma_{1,l^{*}(1)}, \gamma_{2,l^{*}(2)},..,\gamma_{(i-1),l^{*}(i-1)} \}$. Note that the set $\bar{U}_{i}$ contains $L-(i-1)$ CQI metrics which are i.i.d. r.v's. The size of this set is: $|\bar{U}_{i}|=L-(i-1)$. For $i$th SI, the scheduling decision $l^{*}(i)$ is obtained as
$
\gamma_{i,l^{*}(i)}= \texttt{max} \quad \bar{U}_{i}, \quad i=1,2,..,N_t
$
Thus the transmitter selects $N_t$ users using a sequential max-SINR scheduler, and transmits data to these users simultaneously, using a suitable modulation and coding scheme (MCS). Each scheduled user is served at a rate $I_{i}=\log(1+\gamma_{i,l^{*}(i)})$, where $I_{i}$ denotes the mutual information measured at the output of the MMSE receiver of the scheduled user for $i$th SI.
The outage probability for $i$th data stream is given by: $ P_{\texttt{out},i}= P(I_{i} < \log(1+\beta_{i}) $ where $\log(1+\beta_{i})$ is the target outage capacity for $i$th steam.  If we assume that the outage requirement for all steams are equal, we set $\beta_i=\beta$. The outage probability can be expressed in alternative form as:
\begin{eqnarray}\label{Eq:Outage}
P_{\texttt{out},i} = P( \gamma_{i,l^{*}(i)} < \beta)
\end{eqnarray}
where $\gamma_{i,l^{*}(i)}$ is obtained by taking the maxima over $L-(i-1)$ CQI metrics which are i.i.d. r.v's. An expression of this form is encountered in single stream case. Using the outage probability results obtained in the SST case, the outage probability for $i$th data stream is upper-bounded as
\begin{eqnarray}\label{Eq:muexact}
P_{\texttt{out},\texttt{UB}} &=& \left[ 1- e^{-\beta N_t N_{0}} \sum_{k=\bar{k}_0}^{K_0}  \frac{a(k) k!}{\left(\beta+m
\right)^{k+1}} \right ]^{L-(i-1)} \\
&\approx& \left[ 1- e^{-\beta N_t N_{0}} \sum_{k=\bar{k}_0}^{K_0}  \frac{a(k) k!}{\left(\beta+m
\right)^{k+1}} \right ]^{L}, \quad  \texttt{for} \quad L \gg N_t.
\end{eqnarray}
where $\bar{k}_0=KN_t-N_r-1$. This result indicates that, in the limiting case when the number of active users is very large compared to the number of transmit antennas, each stream fully exploits the entire pool of available users for scheduling. The number of active users required to meet a given per stream outage probability is given by
\begin{eqnarray}
L \approx \frac{\ln (P_{\texttt{out},\texttt{UB}})}{\ln \left[ 1- e^{\beta N_t N_{0}} \sum_{k=\bar{k}_0}^{K_0}  \frac{a(k) k!}{\left(\beta+m \right)^{k+1}}
\right ]}
\end{eqnarray}

\textbf{\emph{Remark}}\\
If the successive max SINR scheduler uses  the set $U_{i}$ for scheduling instead of $\bar{U}_{i}$, then the scheduler may assign a variable number of streams to each user.
\section*{Appendix B \\ Capacity scaling laws for real encoding}\label{Appendix3}
\subsection*{Appendix B 1 \\Max-SINR Scheduling based on Post-processing SINR of WL-MMSE}\label{Appendix3:sch}
Although, the TOP analysis for real-encoding case exhibits certain similarities compared to complex case, the performance differs in a significant manner. The following analysis exposes the key differences. In this case, the receiver weighs and combines the real and imaginary parts of the multi-antenna received signal samples using a WL MMSE filter $\tilde{\bw}_{l}$ to produce a
decision variable $z_{l}(k)=\tilde{\bw}_{l} \tilde{\by}_{l}(k)$, where $\tilde{\bw}_{l}=\sqrt{S}\tilde{\bh}^{\dagger}_{l} \bar{\bR}^{-1}_{l}$ and $ \bar{\bR}_{l} =  \sum_{i=1}^{K-1} I_0 \tilde{\bg}_{i,l} {\tilde{\bg}^{\dagger}}_{i,l} + \frac{N_{0}}{2} \bI$ is the WL NICM. The SINR at the output of the WL MMSE receiver is given by: $ \tilde{\gamma}_{l} = S \tilde{\bh}^{\dagger}_{l} \bar{\bR}^{-1}_{l}\tilde{\bh}_{l}$. Let $\tilde{r}_l$ denote the rank of the WL ICM
defined as: $\tilde{\bR}_{l}=\sum_{i=1}^{K-1} I_0 \tilde{\bg}_{i,l} {\tilde{\bg}^{\dagger}}_{i,l}$. Following the approach for conventional case, the post-SINR can be expressed as: $\tilde{\gamma}_{l} =S_{l} \tilde{\bh}^{\dagger}_{l} \bar{\bR}^{-1}_{l}\tilde{\bh}_{l}$ which simplifies to
\begin{eqnarray}
\tilde{\gamma}_{l} = S  \sum_{p=1}^{\tilde{r}_{l}}
         \frac{|\tilde{\omega}_{l,p}|^2}{\tilde{\lambda}_{l,p}+\frac{N_{0}}{2}}+S\sum_{p=\tilde{r}_{l}+1}^{2N_r} \frac{2|\tilde{\omega}_{l,p}|^2}{N_{0}}, \quad l=1,2,..,L
\end{eqnarray}
where $\tilde{\omega}_{l}=[\tilde{\omega}_{l,1}, \tilde{\omega}_{l,2},..,\tilde{\omega}_{l,2N_r}]^{Tr}$ is a real-valued vector that has same
distribution as $\tilde{ \bh}_{l}$. When $ K-1 < 2N_r$, the WL ICM becomes rank deficient and therefore the
receiver at each user can suppress up to $2N_r-1$ interferers fully. In the opposite case when $K-1 \ge 2N_r$, the WL ICM has full rank. As in case of complex-valued signaling, we consider the TOP analysis only for $K-1 \ge 2N_r$. For this case, the post-SINR of WL MMSE takes the form
\begin{eqnarray}
\tilde{\gamma}_{l} =S \sum_{p=1}^{2 N_r} \frac{|\tilde{\omega}_{l,p}|^2}{\tilde{\lambda}_{l,p}+\frac{N_{0}}{2}}
 \ge  S \frac{|\tilde{\omega}_{l,2 N_r}|^2}{\tilde{\lambda}_{l,2 N_r}+\frac{N_{0}}{2}}.
\end{eqnarray}
Let $\hat{\gamma}_{l} = S \frac{|\tilde{\omega}_{l,2 N_r}|^2}{\tilde{\lambda}_{l,2 N_r}+\frac{N_{0}}{2}}$. If we assume that each user reports the SINR
$\hat{\gamma}_{l}$ instead of actual SINR, the TOP can upper bounded as
\begin{eqnarray}
P_{\texttt{out},\texttt{UB,Real}} = P( \texttt{max} \left(\hat{\gamma}_{1}, \hat{\gamma}_{2},..,\hat{\gamma}_{L} \right) < \beta) = \hat{F}^{L}(\beta) \label{Eq:PAMoutage}
\end{eqnarray}
where $\hat{F}(\beta)=P(\hat{\gamma}_{l}< \beta)$ and,
$
\hat{F}(\beta) =P\left(S  \frac{|\tilde{\omega}_{l,2N_r}|^2}{\tilde{\lambda}_{l,2N_r}+\frac{N_{0}}{2}} < \beta \right)
$.
Let $\tilde{\lambda}_{m}=\tilde{\lambda}_{l,2N_r}$ denote the minimum eigenvalue of the WL ICM, and $\hat{x}=|\tilde{\omega}_{l,2N_r}|^2$, where omitted the dependency on index $l$. Since $\tilde{\omega}_{l,2N_r} \sim \textsf{N}(0,\frac{1}{2})$, the pdf of $\hat{x}$ is given by
\begin{eqnarray}
p(\hat{x}) = \frac{1}{\sqrt{\pi \hat{x}}}  e^{-\hat{x}}.  \label{pdfrealx}
\end{eqnarray}
This pdf differs from the case of complex encoding where we deal with exponential distribution.
\subsubsection*{Appendix B 1.1 \\ pdf of minimum eigenvalue of a real Wishart matrix}
Let us consider the WL ICM: $\tilde{\bR}_{l}=\sum_{i=1}^{K-1} I_0 \tilde{\bg}_{i,l} {\tilde{\bg}^{\dagger}}_{i,l}$ where $\tilde{\bg}_{i,l} $ is a real-valued i.i.d.
Gaussian random vector: $\textsf{N}(0, \frac{1}{2}\bI)$. This matrix is called a real Wishart matrix \cite{book:Murihead}, denoted as: $\bW_{n}(m, \frac{I_0}{2} \bI)$, $n\ge m$ where
$m=\texttt{min}(2N_r,K-1)$ and $n=\texttt{max}(2N_r, K-1)$. Since we are considering the specific case of $K-1 \ge 2N_r$, we have: $m=2N_r$, $n=K-1$. The joint pdf of the ordered
eigenvalues $\tilde{\lambda}_1,..,\tilde{\lambda}_m, (\tilde{\lambda}_1>\tilde{\lambda}_2>..>\tilde{\lambda}_m>0)$ of $\bW_{m}(n, \frac{I_0}{2} \bI)$, $n\ge m$ is given in \cite{book:Murihead}. In \cite{Edelman}, the pdf of MEV is expressed for the special case of $I_{0}=2$. The pdf for the general case is obtained by using a transformation: $\lambda \rightarrow
\frac{2\lambda}{I_0}$. For even values of $K$, the pdf takes the form
\begin{eqnarray}
p(\tilde{\lambda}_m)=  \frac{1}{I_0} e^{\frac{-m \tilde{\lambda}_m}{I_0}} \sum_{k=k_0}^{K_0} a(k) \left(\frac{\tilde{\lambda}_m}{I_0}\right)^{k_0+1}
\label{Eq:polynomialform}
\end{eqnarray}
where $k_0=\frac{(K-2N_r-2)}{2}$.  The entries in Table I can be used to obtain the values of $a(k)$ for several combinations of $(n,\frac{K}{2}+1)$. Note that for even values of $K$, the pdf of the MEV of a real Wishart matrix has the same form as that of a complex Wishart matrix. However, for odd values of $K$, the pdf has a remarkably different form. For the special case of
$K=2N_r+1$, the pdf is given by
\begin{eqnarray}
 p(\tilde{\lambda}_m) &=& \Gamma\left(\frac{m+1}{2} \right) \frac{m }{\sqrt{\pi I_0 \tilde{\lambda}_m}} e^{\frac{-m \tilde{\lambda}_m}{I_0}}
 U\left(\frac{m-1}{2},\frac{-1}{2},\frac{\tilde{\lambda}_m}{I_0} \right) \label{Eq:Tricomi}
\end{eqnarray}
where the Tricomi function $U(a,b,z)$ is the confluent hypergeometric function
\begin{eqnarray}
 U(a,b,z)= \frac{1}{\Gamma(a)}\int_{t=0}^{\infty} e^{-zt} t^{a-1} (1+t)^{b-a-1} \quad dt, \quad \Re(a) >0.  \label{Eq:Tricomi1}
\end{eqnarray}
where $\Gamma(x)= \int_{0}^{\infty}t^{x-1} e^{-t}\, dt \, x>0$ is the gamma function and $U\left(a,b,0 \right)=\frac{\Gamma\left(1-b\right)}{\Gamma\left(a-b+1\right)}$. For other values of $K$ taking odd values, the pdf can be obtained using the recursive formula given in \cite{Edelman}.
\subsection*{Appendix B 2 \\ TOP with real-valued encoding}\label{Appendix3:TOPreal}
\begin{figure}
    \centerline{
        \epsfig{figure=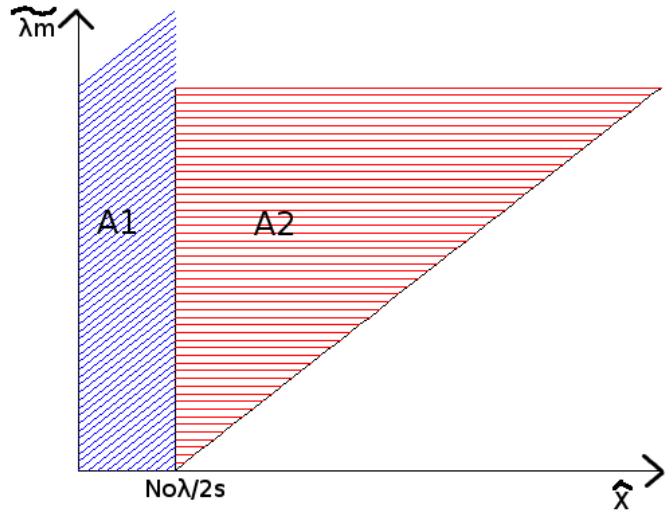, width=10.0cm}
    }
        \vspace*{0.0cm}  \caption{Area under integration} \label{fig:00}

\end{figure}

First, we shall derive an exact expression for the TOP for even values of $K$. $\hat{F}(\lambda)$ is evaluated as
\begin{eqnarray}
\hat{F}(\beta) = P\left( S \frac{\hat{x}}{\tilde{\lambda}_{m}+\frac{N_{0}}{2}} < \beta \right) = P\left( \tilde{\lambda}_{m} >
\frac{S\hat{x}}{\beta}-\frac{N_{0}}{2}  \right). \label{Eq:outagelbwla}
\end{eqnarray}
The expression (\ref{Eq:outagelbwla}) is evaluated by integrating the joint pdf $p(\tilde{\lambda}_{m}, \hat{x})=p(\tilde{\lambda}_{m})p(\hat{x})$ over the shaded area shown in Fig \ref{fig:00}. The area under the region A1 is given by
\begin{eqnarray}
A1 = \int_{\hat{x}=0}^{\frac{\beta N_{0}}{2S}} \frac{1}{\sqrt{\pi} \sqrt{\hat{x}}} e^{-\hat{x}} \quad d \hat{x} \int_{\tilde{\lambda}_{m}=0}^{\infty } p(\tilde{\lambda}_{m}) \quad
d  \tilde{\lambda}_{m}
=1-2Q\left(\sqrt{\frac{\beta N_{0}}{S}} \right). \label{Eq:outagelbwlb}
\end{eqnarray}
A change of variable $ \hat{x}=\frac{y^2}{2}$ is used arrive at the result and $Q(a) \cong \frac{1}{\sqrt{2 \pi}} \int_{a}^{\infty} e^{-\frac{a^2}{2}}$. For even values of $K$, the area A2 is evaluated
as
\begin{eqnarray}
A2 &=& \int_{\hat{x}=\frac{\beta N_{0}}{2S}}^{\infty} \left[ \int_{\tilde{\lambda}_{m}=\frac{S\hat{x}}{\beta}-\frac{N_{0}}{2}}^{\infty } p(\tilde{\lambda}_{m}) \quad
d  \tilde{\lambda}_{m} \right]p(\hat{x}) \quad d \hat{x}. \label{Eq:outageWLcase1}
\end{eqnarray}
The integral is evaluated in the Appendix C and the result is given in (\ref{Eq:outageWLcase1c}). The TOP is determined as
\begin{eqnarray}
P_{\texttt{out},\texttt{UB,Real}} &=& (A1+A2)^{L} \quad K  \quad \texttt{even}. \label{Eq:outagewlf}
\end{eqnarray}
\subsection*{Appendix B 3 \\ Further Approximations}\label{Appendix3:approx}
The TOP expression (\ref{Eq:outagewlf}) allows fast and easy numerical computation but it is not in a form convenient to illustrate the trade-off between the number of required users and associated interference suppression effects. We present an alternative result using certain approximations. This approach is applicable to both even and odd
values of $K$. To this end, we evaluate $\hat{F}(\lambda)$ as
\begin{eqnarray}
\hat{F}(\beta)=P\left(S\frac{\hat{x}}{\tilde{\lambda}_{m}+\frac{ N_{0}}{2}} < \beta \right) = \int_{\tilde{\lambda}_m=0}^{\infty} P \left(\hat{x} < \beta
\frac{(\tilde{\lambda}_{m}+\frac{ N_{0}}{2})}{S}\right) p(\tilde{\lambda}_m) \quad  d \tilde{\lambda}_m. \label{Eq:Altern}
\end{eqnarray}
Consider
\begin{eqnarray}
P \left(\hat{x} < \beta \frac{(\tilde{\lambda}_{m}+\frac{ N_{0}}{2})}{S}\right) = \int_{\hat{x}=0}^{\beta \frac{(\tilde{\lambda}_{m}+\frac{ N_{0}}{2})}{S}}
\frac{1}{\sqrt{\pi \hat{x}}} e^{-\hat{x}} \quad d\hat{x}
=1-2Q \left(\sqrt{2\beta \frac{(\tilde{\lambda}_{m}+\frac{ N_{0}}{2})}{S}}\right).
\end{eqnarray}
This expression is not suitable for closed-form evaluation of  (\ref{Eq:Altern}). To arrive at simple expression, the $Q$-function is approximated as a sum of exponentials
as: $Q(x) \approx \frac{1}{12} e^{-\frac{x^2}{2}} + \frac{1}{4} e^{-\frac{2x^2}{3}}$. This is tight approximation for a wide range of values of $x$ \cite{Chiani:July03}.
Using this
\begin{eqnarray}
P \left(\hat{x} < \beta \frac{(\tilde{\lambda}_{m}+\frac{ N_{0}}{2})}{S}\right) & \approx & 1-2 \sum_{i=1}^{2}K_i e^{-2c_i \beta \frac{(\tilde{\lambda}_{m}+\frac{
N_{0}}{2})}{S}}\label{Eq:Qbased}
\end{eqnarray}
where $c_1=\frac{1}{2}, c_2=\frac{2}{3}, K_1=\frac{1}{12}, K_2=\frac{1}{4}$. Substituting (\ref{Eq:Qbased}) in (\ref{Eq:Altern}), we get
\begin{eqnarray}
\hat{F}(\beta)& \approx &  1-2 \sum_{i=1}^{2}K_i \int_{\tilde{\lambda}_m=0}^{\infty} e^{-2c_i \beta \frac{(\tilde{\lambda}_{m}+\frac{ N_{0}}{2})}{S}}
p(\tilde{\lambda}_m) \quad  d \tilde{\lambda}_m.  \label{Eq:Alternapprox}
\end{eqnarray}
\subsection*{Appendix B 4 \\ Approximations for even values of $K$}\label{Appendix3:approx2}
Substituting the pdf $p(\tilde{\lambda}_m)$ given by (\ref{Eq:polynomialform}) for even values of $K$,  we get
\begin{eqnarray}
\hat{F}(\beta)& \approx &  1-2 \sum_{i=1}^{2}K_i e^{-c_i \beta \frac{ N_{0}}{S}} \left[\sum_{k=k_0}^{K_0} a(k) \int_{\tilde{\lambda}_m=0}^{\infty}
\frac{1}{I^{k+1}_0}e^{-\tilde{\lambda}_{m} \left( \frac{(2c_i \beta}{S} +\frac{m}{I_0} \right)} \tilde{\lambda}^{k}_m \quad  d \tilde{\lambda}_m  \right]\\
&=& 1-2 \sum_{i=1}^{2}K_i e^{-c_i \beta \frac{ N_{0}}{S}} \left[\sum_{k=k_0}^{K_0} a(k) \frac{k !}{\left( \frac{(2c_i I_0 \beta}{S} +m\right)^{k+1}} \right].
\end{eqnarray}
A change of variable $u=\tilde{\lambda}_{m} \left( \frac{(2c_i \beta}{S} +\frac{m}{I_0} \right)$ is made on line 1 to arrive at the result. Using (\ref{Eq:PAMoutage}), the total number of active
users required to meet a given TOP is given by
\begin{eqnarray}
L & \approx & \frac{ \ln(P_{\texttt{out,UB,Real}})}{ \ln\left[1-2 \sum_{i=1}^{2} K_i e^{-c_i \beta \frac{ N_{0}}{S}} \left[ \sum_{k=k_0}^{K_0} a(k) \frac{k!}{\left(
\frac{2c_i I_0 \beta}{S} +m \right)^{k+1}} \right] \right]}\\ & \approx & \frac{ \ln (P^{-1}_{\texttt{out,UB,Real}}) \left( \frac{2c_i I_0 \beta}{S}+m
\right)^{k_0+1}}{2 \sum_{i=1}^{2} K_i e^{-c_i \beta \frac{ N_{0}}{S}} a(k_0) k_0!}.
\end{eqnarray}
To arrive at the result, we assume large values for $\frac{I_0 \beta}{S}$, we invoke the approximation $\ln(1-x) \approx -x$, and retained only the term containing
$k_0$. Substituting, $k_0=\frac{(K-2N_r-2)}{2}$, we get: $L \propto \left( \frac{ I_0 \beta}{S}+m\right)^{\frac{K}{2}-N_r}$.
\subsection*{Appendix B 5 \\Approximations for odd values of $K$}\label{Appendix3:odd}
Evaluation of TOP for odd values of $K$ is considerably more involved. In the following, we provide the TOP expression for the case of $K=2N_r+1$ and for the case of $K=2N_r+3$. Results for the general case of $K$ taking odd values are omitted due to space limitations.
\subsubsection*{Appendix B 5.1 \\Approximations for $K=2N_r+1$}\label{Appendix3:oddcase1}
Substituting (\ref{Eq:Tricomi}) in  (\ref{Eq:Alternapprox}), after simple manipulations we get
\begin{eqnarray}
\hat{F}(\beta) & \approx &  1-2\Gamma\left(\frac{m+1}{2} \right) \frac{m }{ \sqrt{\pi} }
\sum_{i=1}^{2} K_i e^{-c_i \beta \frac{ N_{0}}{S}} \int_{u=0}^{\infty}  \frac{1}{\sqrt{u\left( 2c_i I_0 \frac{\beta}{S} + m\right)}} e^{-u} \times \nonumber \\
& & \quad \quad
U\left(\frac{m-1}{2},\frac{-1}{2},\frac{u}{\left( 2c_i I_0 \frac{\beta}{S} + m\right)} \right) \quad  du. \label{Fevenneqm}
\end{eqnarray}
Note that, for large values of $u$, the integrand takes small values due to the scaling factor $\frac{e^{-u}}{\sqrt{u}}$. Therefore, it is sufficient to consider
the integrand for small to medium values of $u$. At high SNR, $\frac{u}{\left( 2c_i I_0 \frac{\lambda}{S} + m\right)}$ takes very small values. Under this assumption, and using the result: $U\left(a,b,0 \right)=\frac{\Gamma\left(1-b\right)}{\Gamma\left(a-b+1\right)}$, the tricomi function is approximated as
\begin{eqnarray}
U\left(\frac{m-1}{2},\frac{-1}{2},\frac{u}{\left( 2c_i I_0 \frac{\beta}{S} + m\right)} \right) \approx  U\left(\frac{m-1}{2},\frac{-1}{2},0 \right)=
\frac{\Gamma\left(\frac{3}{2}\right)}{\Gamma\left(\frac{m}{2}+1\right)}. \label{Uzeroapprox}
\end{eqnarray}
Substituting (\ref{Uzeroapprox}) in (\ref{Fevenneqm}), after simple manipulations we get
\begin{eqnarray}
\hat{F}(\beta)  \approx 1-2m\Gamma\left(\frac{m+1}{2} \right) \Gamma\left(\frac{3}{2} \right)\sum_{i=1}^{2} \frac{K_i e^{-c_i \beta \frac{ N_{0}}{S}} }{\Gamma\left(\frac{m}{2}+1 \right) \left( \sqrt{2c_i I_0 \frac{\beta}{S}+m} \right)}.
\end{eqnarray}
Using this
\begin{eqnarray}
L \approx  \frac{ \ln (P_{\texttt{out,UB,Real}})}{ \ln\left[1-2m\Gamma\left(\frac{m+1}{2} \right) \Gamma\left(\frac{3}{2} \right)\sum_{i=1}^{2} \frac{K_i e^{-c_i \beta \frac{ N_{0}}{S}} }{\Gamma\left(\frac{m}{2}+1 \right) \left( \sqrt{2c_i I_0 \frac{\beta}{S}+m} \right)}\right]} \approx \frac{ \ln (P_{\texttt{out,UB,Real}}) \sqrt{ I_0 \frac{\beta}{S}}}{ 2m\Gamma\left(\frac{m+1}{2} \right) \Gamma\left(\frac{3}{2} \right)\sum_{i=1}^{2} \frac{K_i e^{-c_i \beta \frac{ N_{0}}{S}} }{\Gamma\left(\frac{m}{2}+1 \right)\sqrt{2c_i}}} \label{Eq:Meven1}
\end{eqnarray}
where we assume $ 2c_i I_0 \frac{\beta}{S} +m$ to take high values and ${ 2c_i I_0 \frac{\beta}{S}} \gg  m $. In this case, $L$ is directly proportional to the square root of $I_0 \frac{\beta}{S}$.
\subsection*{Appendix B 6\\ TOP for real-valued Encoding $K=2N_r+3$}\label{Appendix3:oddcase2}
For $K=2N_r+3$, $p(\tilde{\lambda}_m)$ is given by \cite{Edelman}
\begin{eqnarray}
 p(\tilde{\lambda}_m) &=& \Gamma\left(\frac{m+1}{2} \right) \frac{2}{\sqrt{\pi} I^{\frac{3}{2}}_0} \sqrt{\tilde{\lambda}_m} e^{\frac{-m \tilde{\lambda}_m}{I_0}}
 g(\tilde{\lambda}_m) \label{pdf2}
\end{eqnarray}
where
\begin{eqnarray}
g(\tilde{\lambda}_m) &=& L^{(2)}_{m-1}\left(-\frac{2\tilde{\lambda}_m}{I_0}\right) U\left(\frac{m-1}{2},\frac{-1}{2},\frac{\tilde{\lambda}_m}{I_0} \right) +
\frac{\tilde{\lambda}_m}{I_0} L^{(3)}_{m-2}\left(-\frac{\tilde{\lambda}_m}{I_0}\right) U\left(\frac{m+1}{2},\frac{1}{2},\frac{\tilde{\lambda}_m}{I_0}
\right)\label{pdf2}
\end{eqnarray}
and $L^{(\alpha)}_{p} (-x)= \sum_{q=0}^{p} (p+\alpha)C_{p-q} x^{q}$. Substituting (\ref{pdf2}) in (\ref{Eq:Alternapprox}) we get
\begin{eqnarray}
\hat{F}(\beta) & \approx &  1-2\Gamma\left(\frac{m+1}{2} \right) \frac{2}{ \sqrt{\pi} I^{\frac{3}{2}}_0 } \sum_{i=1}^{2} K_i e^{-c_i \beta \frac{ N_{0}}{S}}
\int_{\tilde{\lambda}_m=0}^{\infty}  \sqrt{\tilde{\lambda}_m} e^{- \tilde{\lambda}_m\left( 2c_i \frac{\beta}{S} + \frac{m}{I_0}\right)}g(\tilde{\lambda}_m) \quad  d
\tilde{\lambda}_m. \label{Eq:outageEven2}
\end{eqnarray}
Consider
\begin{eqnarray}
\frac{1}{I^{\frac{3}{2}}_0}\int_{\tilde{\lambda}_m=0}^{\infty}  \sqrt{\tilde{\lambda}_m} e^{- \tilde{\lambda}_m\left( 2c_i \frac{\beta}{S} +
\frac{m}{I_0}\right)}g(\tilde{\lambda}_m) \quad  d \tilde{\lambda}_m & = &  \frac{1}{\left( 2c_i \frac{I_0 \beta}{S} + m\right)^{\frac{3}{2}}} \int_{u=0}^{\infty}
\sqrt{u} e^{- u} \times  \nonumber\\
& & \quad \quad  g\left(\frac{u}{ 2c_i \frac{ \beta}{S} + \frac{m}{I_0}}\right) \quad  d u \label{Eq:outageEven3}
\end{eqnarray}
where
\begin{eqnarray}
g\left(\frac{u}{ 2c_i \frac{ \beta}{S} + \frac{m}{I_0}}\right) &=& L^{(2)}_{m-1}\left(-\frac{2u}{ 2c_i \frac{ I_0 \beta}{S} + m}\right)
U\left(\frac{m-1}{2},\frac{-1}{2}, \frac{u}{ 2c_i \frac{ I_0 \beta}{S} + m}\right) \\ & & \quad + \left(\frac{u}{ 2c_i \frac{ I_0 \beta}{S} + m}\right)
L^{(3)}_{m-2}\left(-\frac{2u}{ 2c_i \frac{ I_0 \beta}{S} + m}\right) U\left(\frac{m+1}{2},\frac{1}{2},\frac{u}{ 2c_i \frac{ I_0 \beta}{S} + m} \right)\\
& \approx &L^{(2)}_{m-1}(0)
U\left(\frac{m-1}{2},\frac{-1}{2}, 0\right).
\end{eqnarray}
The approximation in second line holds for high values of $2c_i \frac{ \beta}{S} + \frac{m}{I_0}$. Since, $ U\left(\frac{m-1}{2},\frac{-1}{2},
0\right)=\frac{\Gamma\left(\frac{3}{2}\right)}{\Gamma\left(\frac{m}{2}+1\right)}$, and $ L^{(2)}_{m-1}(0)
= (m+1)C_{m-1}$, we get
\begin{eqnarray}
g\left(\frac{u}{ 2c_i \frac{ \beta}{S} + \frac{m}{I_0}}\right) &\approx& (m+1)C_{m-1} \frac{\Gamma\left(\frac{3}{2}\right)}{\Gamma\left(\frac{m}{2}+1\right)}.
\label{Eq:outageEven4}
\end{eqnarray}
Substituting (\ref{Eq:outageEven4}), and (\ref{Eq:outageEven3}) in (\ref{Eq:outageEven2}), we get
\begin{eqnarray}
\hat{F}(\beta) & \approx &  1- (m+1)C_{m-1} 4\Gamma\left(\frac{3}{2}\right) \sum_{i=1}^{2} \frac{K_i
e^{-c_i \beta \frac{ N_{0}}{S}}}{\left( 2c_i \frac{I_0 \beta}{S} + m\right)^{\frac{3}{2}}}  \int_{u=0}^{\infty}  \frac{1}{\sqrt{\pi}}\sqrt{u} e^{- u} \quad du \label{Eq:Meven2}\\
&=& 1- (m+1)C_{m-1} \sqrt{\pi} \sum_{i=1}^{2} \frac{K_i
e^{-c_i \beta \frac{ N_{0}}{S}}}{\left( 2c_i \frac{I_0 \beta}{S} + m\right)^{\frac{3}{2}}}.
\end{eqnarray}
The result on second line is due to $ \int_{u=0}^{\infty}  \frac{2}{\sqrt{\pi}}\sqrt{u} e^{- u} \quad du =1$, and $\Gamma\left(\frac{3}{2}\right)=\frac{\sqrt{\pi}}{2}$.

Now combining the results for odd and even cases, we get: $L \propto \left( \frac{ I_0 \beta}{S}+m\right)^{\frac{K}{2}-N_r}$. For $\beta=\frac{S}{N_0}$, for $S=I_0$ and when $\frac{ I_0 \beta}{S} > m$, we have: $L \propto SNR^{\frac{K}{2}-N_r}$. The sum of outage capacities of $K$ transmitters employing real-valued encoding is given by
\begin{eqnarray}
C_{\texttt{sum}, \texttt{Real}} &=& \frac{K}{2}\log(1+\beta)
\end{eqnarray}
This expression holds when the number of users is sufficiently high.

\section*{Appendix C}\label{sec:Appendix}
First evaluate area
$
A2a= \int_{\tilde{\lambda}_{m}=\frac{S\hat{x}}{\beta }-\frac{N_{0}}{2}}^{\infty } p(\tilde{\lambda}_{m}) \quad d  \tilde{\lambda}_{m}
  = \sum_{k=k_0}^{K} \frac{a(k)}{m^{(k+1)}}    \int_{t=t_0}^{\infty } e^{-t} t^k \quad d  t
$
where a change of variable $t=\frac{m \hat{\lambda}_m}{I_0}$ is made on line 2 and we define $t_0=m \left(\frac{ S\hat{x}}{I_0 \beta }-\frac{N_{0}}{2 I_0} \right)$.
Using integration by parts we have: $ \int_{t=t_0}^{\infty} t^k e^{-t} \quad dt = \sum_{p=0}^{k} e^{-t_0} {t_0}^{(k-p)} \frac{k!}{(k-p)!}$. Using this
\begin{eqnarray}
A2a  &=& \sum_{k=k_0}^{K} \frac{a(k)}{m^{(k+1)}} \left[ \sum_{p=0}^{k} e^{-\left(
\frac{m S\hat{x}}{I_0 \beta}-\frac{mN_{0}}{2 I_0} \right)} m^{(k-p)} \left[  \sum_{r=0}^{k-p} (k-p)C_{r} \left(\frac{-N_{0}}{2 I_0} \right)^{r} \left(
\frac{S\hat{x}}{I_0 \beta} \right)^{(k-p-r)} \right] \right]
\end{eqnarray}
where binomial expansion of $\left(  \frac{S\hat{x}}{I_0 \beta}-\frac{N_{0}}{2 I_0} \right)^{(k-p)}$ is used. Substituting A2a into
(\ref{Eq:outageWLcase1}), we have
\begin{eqnarray}
A2 &=& \int_{\hat{x}=\frac{\beta N_{0}}{2S}}^{\infty}  \sum_{k=k_0}^{K} \frac{a(k)}{m^{(k+1)}}  \sum_{p=0}^{k} e^{-\left( \frac{m S\hat{x}}{I_0
\beta}-\frac{mN_{0}}{2 I_0} \right)} \left(\frac{m}{I_0}\right)^{(k-p)}  \sum_{r=0}^{k-p} (k-p)C_{r} \left(\frac{-N_{0}}{2} \right)^{r} \left( \frac{S\hat{x}}{\beta}
\right)^{(k-p-r)} \frac{1}{\sqrt{\pi} \sqrt{\hat{x}}} e^{-\hat{x}} \quad d\hat{x}  \nonumber \\
&=& \frac{1}{\sqrt{\pi}} e^{\frac{mN_{0}}{2 I_0}}   \sum_{k=k_0}^{K} \frac{a(k)}{m^{(k+1)}} \sum_{p=0}^{k} \left(\frac{m}{I_0}\right)^{(k-p)}  \sum_{r=0}^{k-p}
(k-p)C_{r} \left(\frac{-N_{0}}{2} \right)^{r} \left( \frac{S}{\beta} \right)^{(k-p-r)} \frac{D}{\left( \frac{m S}{I_0 \beta} +1 \right)^{k-p-r+\frac{1}{2}}}
\label{Eq:outageWLcase1c}
\end{eqnarray}
where $D=\frac{1}{\sqrt{\pi}}\int_{\hat{x}=\frac{\beta N_{0}}{2S} \left({\frac{m S}{I_0 \beta} +1} \right)}^{\infty} u^{\left(k-p-r-\frac{1}{2}\right)}e^{-u}  du $. Using integration by parts, this expression can be represented in terms of Q-function which is suitable for numerical calculation.

\begin{table}\label{Table-1}
\centering \caption{Coefficients $a(k)$}
\begin{tabular}{|c|c|}
  \hline
$m$, $n$ & $a(k)$ \\
  \hline
(2,2) & [2]  \\
  (2,3) & [0 4/3 8/3]  \\
(2,4) &  [ 0 0 24 16 4]/15\\
  \hline
  (4,4) & [4]  \\
  (4,5) & [0 120 180 72 8 ]/15  \\
(4,6) &  [0 0 40320 80640 72000 33600 8640 1152 64]/6300\\
    \hline
\end{tabular}
\end{table}

\begin{figure}
    \centerline{
        \epsfig{figure=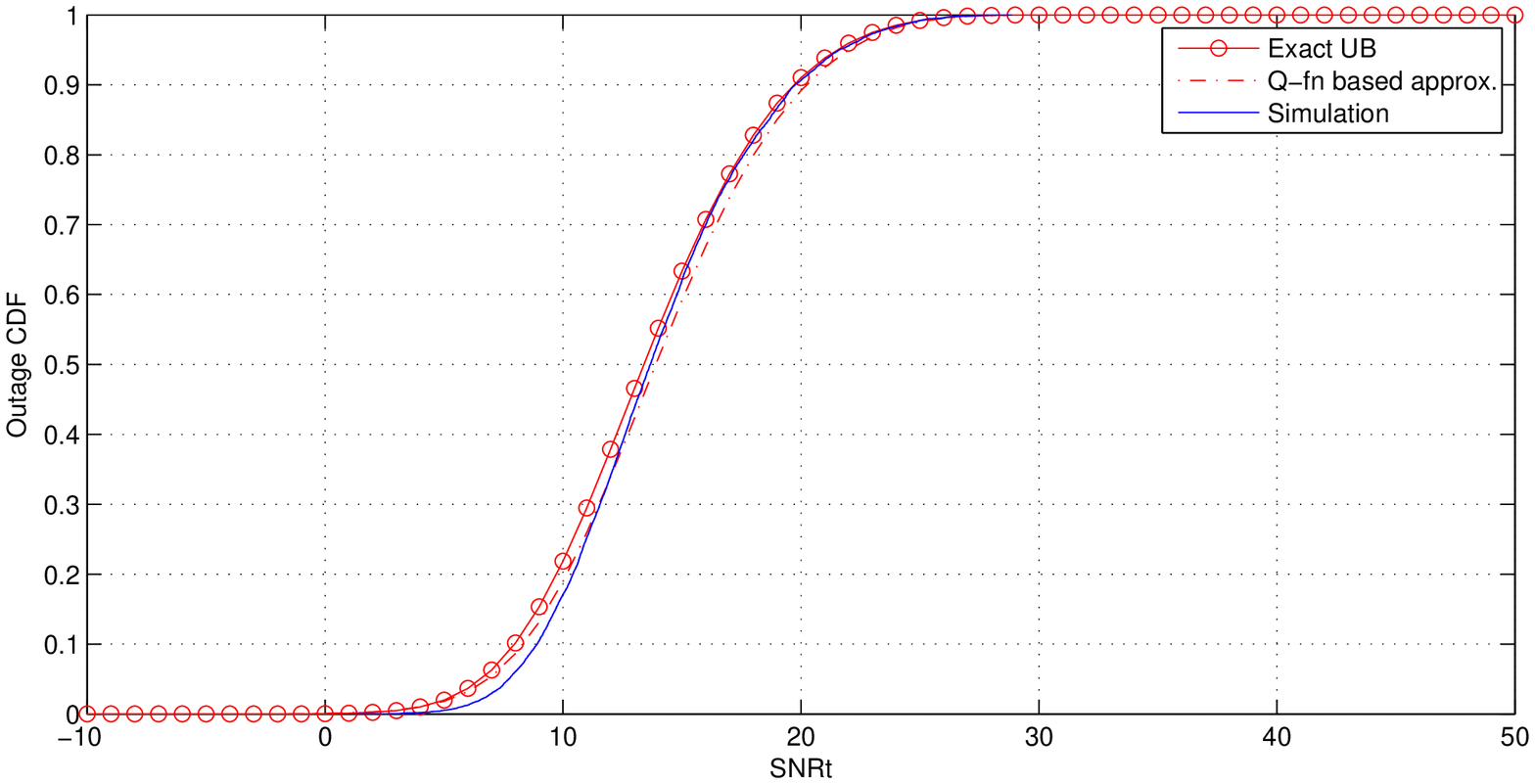, height=10cm, width=15.0cm}
    }
        \vspace*{0.0cm}  \caption{TOP for real-valued encoding,  $N_r=2$, $K=6$, $L=10$, SNR=20 dB} \label{fig:1}

\end{figure}

\begin{figure}
    \centerline{
        \epsfig{figure=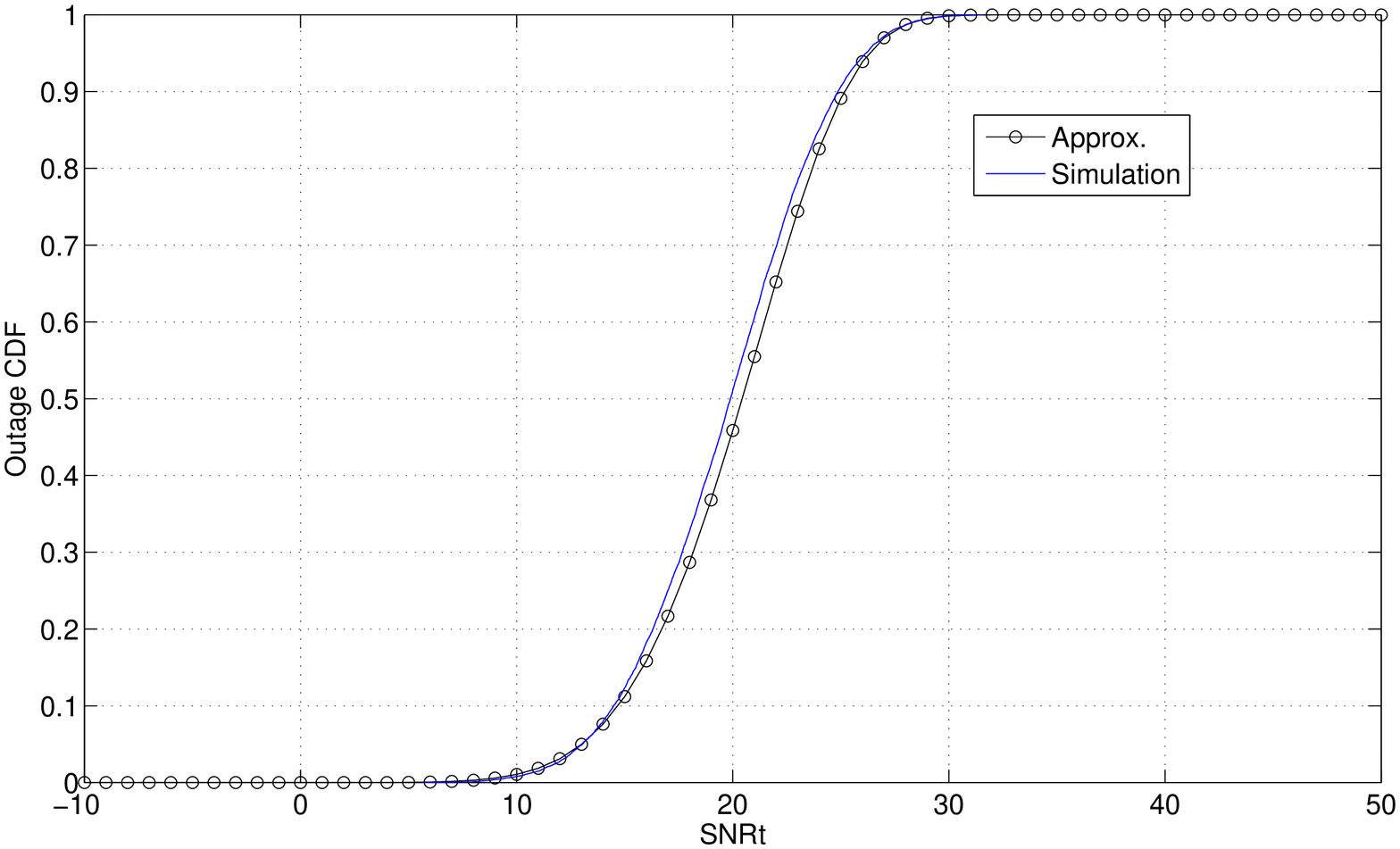, height=10cm, width=15.0cm}
    }
        \vspace*{0.0cm}  \caption{TOP for real-valued encoding,  $N_r=2$, $K=5$, $L=10$, SNR=20 dB} \label{fig:1a}

\end{figure}

%

%

\begin{table}\label{Table-2}
\centering \caption{\texttt{Mean Sum Capacity in bits/s/Hz for $N_r=1$}}
\begin{tabular}{|c|c|c|c|c|}
\hline
\texttt{Mode} & $L=10$, $SNR=5$ & $L=10$ $SNR=30$ & $L=50$, $SNR=5$ & $L=50$, $SNR=30$\\
 \hline
\texttt{ SST Complex}, $K=3$ & 4.75 & 6.5 & 6.9 & 9.5\\
\hline
\texttt{ SST Complex}, $K=4$ & 4.75 & 5.5 & \textbf{7.1} & 8.5\\
\hline
\texttt{SST Real},    $K=3$  & 4.7  &  \textbf{11}  & 6.0 & 12.5\\
\hline
\texttt{SST Real},    $K=4$  & \textbf{8.35} & 7.9 & 6.9 & \textbf{14.9}\\
\hline
\end{tabular}
\end{table}

\begin{table}\label{Table-3}
\centering \caption{\texttt{Mean Sum Capacity in bits/s/Hz for $N_r=2$}}
\begin{tabular}{|c|c|c|c|c|}
\hline
\texttt{Mode} & $L=50$, $SNR=5$ & $L=50$, $SNR=20$\\
 \hline
\texttt{ SST Complex}, $K=3$ & 9.97 & 21.4\\
\hline
\texttt{ SST Complex}, $K=4$ & \textbf{10.9} & 15.84 \\
\hline
\texttt{ SST Complex}, $K=5$ & \textbf{10.9} & 13.66\\
\hline
\texttt{SST Real},    $K=3$  & 7.34 & 19.67\\
\hline
\texttt{SST Real},    $K=4$  & 9.2 & 25.14\\
\hline
\texttt{SST Real},    $K=5$  & 10.5 & \textbf{25.97}\\
\hline
\end{tabular}
\end{table}

%

\end{document}